\documentclass[final,authoryear,5p,times,twocolumn]{elsarticle}
\usepackage{graphicx, subfig}
\usepackage{amsmath, amsthm, amssymb}
\usepackage{rotating,array}

\usepackage{flafter}

\begin{document}

\title{Dynamics of Rotationally Fissioned Asteroids: \\
Source of Observed Small Asteroid Systems}

\author{Seth A. Jacobson\corref{cor1}}
\ead{seth.jacobson@colorado.edu}
\address{Department of Astrophysical and Planetary Sciences \\
University of Colorado, Boulder, CO 80309 \\
Phone: 303-492-7826}

\author{Daniel J. Scheeres}
\ead{scheeres@colorado.edu}
\address{Department of Aerospace and Engineering Sciences \\
University of Colorado, Boulder, CO 80309 }

\cortext[cor1]{Corresponding author}

\maketitle

Abstract:

We present a model of near-Earth asteroid (NEA) rotational fission and ensuing dynamics that describes the creation of synchronous binaries and all other observed NEA systems including: doubly synchronous binaries, high-$e$ binaries, ternary systems, and contact binaries. Our model only presupposes the Yarkovsky-O'Keefe-Radzievskii-Paddack (YORP) effect, ``rubble pile'' asteroid geophysics, and gravitational interactions. The YORP effect torques a ``rubble pile'' asteroid until the asteroid reaches its fission spin limit and the components enter orbit about each other \citep{Scheeres2007b}. Non-spherical gravitational potentials couple the spin states to the orbit state and chaotically drive the system towards the observed asteroid classes along two evolutionary tracks primarily distinguished by mass ratio. Related to this is a new binary process termed secondary fission--the secondary asteroid of the binary system is rotationally accelerated via gravitational torques until it fissions, thus creating a chaotic ternary system. The initially chaotic binary can be stabilized to create a synchronous binary by components of the fissioned secondary asteroid impacting the primary asteroid, solar gravitational perturbations, and mutual body tides. These results emphasize the importance of the initial component size distribution and configuration within the parent asteroid. NEAs may go through multiple binary cycles and many YORP-induced rotational fissions during their approximately 10 Myr lifetime in the inner solar system. Rotational fission and the ensuing dynamics are responsible for all NEA systems including the most commonly observed synchronous binaries.

Keywords: 

Asteroids, dynamics \sep Asteroids, rotation \sep Near-Earth objects \sep Satellites, formation \sep Satellites of asteroids

\newpage

\section{Introduction}
\label{sec:introduction}

The Near-Earth asteroid (NEA) population has a relatively large amount of data compared to other small body populations, including detailed information on asteroid figures and binary structure, often made possible through the combination of lightcurve and radar techniques. Observers have discovered a wide and complex set of asteroid systems that before this study have not been tied together into a coherent theory. The emergence of radiative forces as a major evolutionary mechanism for small bodies, in particular for NEA systems due to their small size and proximity to the Sun,  makes the development of such a theory possible.

[Figure 1]

A simple model of NEA evolution constructed from the Yarkovsky-O'Keefe-Radzievskii-Paddack (YORP) and binary YORP (BYORP) effects, ``rubble pile'' asteroid geophysics, and gravitational interactions can incorporate all of the diverse observed asteroid classes as shown in Figure 1: synchronous binaries, doubly synchronous binaries, contact binaries, asteroid pairs, re-shaped asteroids, and stable ternaries.

\subsection{Observed NEA Classes}
\label{sec:observedclasses}

[Table 1]

Binary asteroid systems comprise a significant fraction ($15 \pm 4\%$) of the NEA population~\citep{Margot2002, Pravec2006} and include all compositional classes and size scales~\citep{Pravec2007}. Most of these systems are synchronous binaries--the orbital and secondary spin periods are equal, but the primary has a faster spin rate. Observed synchronous systems have mass ratios~$\lesssim0.2$, a system semi-major axis of $1.5$ to $3$ primary diameters, and a possibly elongated secondary and a nearly spherical primary with a distinctive shape characterized by an equatorial bulge~\citep{Pravec2006, Pravec2007}. The system has a positive free energy, but the tidally locked secondary inhibits disruption. Migration to the inner solar system from the main belt as binaries, binary creation via collision amongst NEAs, and binary creation via tidal disruption from close planetary flybys are not efficient enough mechanisms to create this population nor match the observed synchronous binary properties~\citep{Margot2002, Walsh2008a}. Several theories attempt to explain this binary population by a YORP-induced rotational fission process, but do not capture all properties of synchronous binaries and do not predict the other NEA systems that are seen~\citep{Scheeres2007b, Scheeres2009a, Scheeres2009b, Walsh2008b}. 

The~\citet{Walsh2008b} theory requires rotational fission induced ``landslides'' that re-shape the primary, then enter into orbit. Secondaries are built from collections of ``landslide'' material in orbit after many ``landslide'' events, and consequently YORP cycles--the length of the process to rotationally accelerate an asteroid to spin fission from its current state under the YORP effect. However, we will show that material entering orbit via rotational fission will almost always escape on timescales much shorter than a YORP cycle. Furthermore,~\citet{Holsapple2010} using continuum approximations of granular theory finds that mass loss would not occur at the equator of small, critically spinning asteroids, but that their shapes would deform, elongating the body until interior failure. These deformations in the shape of the body allow YORP to continue to increase the angular momentum without significant changes to the spin rate, even slightly decreasing the spin rate in some cases.~\citet{Scheeres2011} reports a similar analytic finding that when cohesive theory is considered, failure will most likely occur along interior planes. The analytic theory in~\citet{Scheeres2009b} describes the first stage of the model proposed herein, where a chaotic binary system is immediately formed from the rotational fissioning of a ``rubble pile.'' A rotational fission model related to the one proposed in this work has been implicated in the formation of asteroid pairs  \citep{Pravec2010}--two asteroids with heliocentric orbits that in the recent past ($\lesssim 10^6$ yrs) intersect deep within the other's Hill radius and with small relative speeds  \citep{Vokrouhlicky2008}. Asteroid pairs are observed in the Main Asteroid Belt with similar sizes to NEAs, but they have not been observed in the NEA population. The theory outlined in this paper predicts them; asteroid pairs are difficult to detect in the NEA population because their orbits are rapidly perturbed and smaller initial asteroids fission into even smaller secondaries for the same mass ratios as the Main Belt asteroid population.

Other observed distinct dynamical and morphological classes include doubly synchronous binaries, high-$e$ binaries, ternaries, contact binaries, and re-shaped asteroids. We describe each in turn. Doubly synchronous binaries: all spin rotation periods are equivalent to the orbital revolution period. They also have high mass ratios $\gtrsim0.2$ and a system semi-major axis of $2$ to $8$ primary diameters  \citep{Pravec2007}. These systems are difficult to detect because of an observational bias in light curve data; doubly synchronous systems and elongated single objects appear similarly. Contact binaries: bimodally-shaped asteroids observed as two similar-sized components resting on each other, which implies a formation mechanism that brings the two components together very gently. Contact binaries comprise a significant fraction ($>9\%$) of the NEA population  \citep{Benner2006}. High-$e$ binaries: low mass ratio binary systems distinct from the synchronous binaries, because they are asynchronous and have high eccentricities  \citep{Taylor2008}. Ternary systems: large primary orbited by two smaller satellites. The primaries are spinning faster than the orbital rates and the mass ratio is low ($<0.1$)  \citep{Brozovic2009}. Re-shaped asteroids: single bodies similar to the primaries of the synchronous binary class--an oblate shaped figure with an equatorial bulge. For reference, examples of each NEA class are given in Table 1. 

\subsection{Motivation}
\label{sec:motivation}

A collisionally evolved asteroid can be modeled as a ``rubble pile"--a collection of gravitationally bound boulders with a distribution of size scales and very little tensile strength between them~\citep{Michel2001, Richardson2005, Tanga2009}. ``Rubble pile'' morphology has been closely examined by the Hayabusa mission to Itokawa, as shown in Fig. 2, which has no obvious impact craters and appears as collection of shattered fragments of different size scales~\citep{Fujiwara2006}. Mass and volume measurements from the NEAR Shoemaker flyby of Mathilde~\citep{Yeomans1997} and radar observations of 1999 KW$_4$~\citep{Ostro2006} determine mean densities that are lower than their constitutive elements, which is evidence of voids and cracks in the structures of these bodies. Asteroids with diameters larger than $\sim200$ m rarely spin with periods less than $\sim2.2$ hours, which corresponds with the critical disruption spin rate of self-gravitating, ``rubble pile'' bodies~\citep{Pravec2007B}. Both theoretical modeling and direct observation indicate that asteroids within a size range of $\sim100$ m to $\sim10$ km have ``rubble pile'' geophysics.

The details of the rotational fission process determine the initial conditions for the binary system. The torque from the YORP effect will increase the centrifugal accelerations acting on each ``rubble pile'' component. There is a specific spin rate at which each component of the body will go into orbit about the rest determined by the largest separation distance of the mass centers of the fissioned component and the remainder of body~\citep{Scheeres2009a}. The smaller component is now the secondary, and the remainder is the primary, both in orbit about each other. The motivation for this study was to determine what happens dynamically after a rotational fission event.

This paper will utilize some important concepts throughout that will be briefly introduced here and further defined later. The mass ratio is defined as the secondary mass divided by the primary mass. The primary of a binary system is always larger than the secondary, so the mass ratio is a number between $0$ and $1$. Secondary fission is rotational fission of the secondary induced via spin-orbit coupling and occurring during the chaotic binary stage of low mass ratio evolution creating chaotic ternaries. Ternary systems have three members. The components in decreasing mass are labeled primary, secondary, and tertiary, however the two smaller members are referred collectively as secondaries. Secondaries may escape the system if the system has a positive free energy. The free energy of an asteroid system is the sum of the kinetic and mutual potential energies of the system (both rotational and translational) neglecting the self-potentials of each body. 

\section{Methods}
\label{sec:methods}

\subsection{Initial Conditions}
\label{sec:initial conditions}

As the rotation rate of an asteroid increases due to the YORP effect, the asteroid will go through a series of reconfiguration events. These events may range from small-scale ``rock flows'' to significant restructuring of the principal coherent components. A ``rubble pile'' will have a specific set of local minimum energy configurations that the coherent components will settle into \citep{Scheeres2009b}. As the spin rate of the asteroid increases due to the YORP effect, the system will act to arrange itself into the global minimum energy configuration. The largest distance between two mass centers, which encompass all of the mass, determines how the system will fission, since systems in this minimum energy equilibrium will undergo rotational fission at lower spin rates than any other configuration \citep{Scheeres2009b}. The model assumes that the system is in this minimum energy configuration and the two components can be represented by tri-axial ellipsoids resting so that contact is along the largest axis of each body.

If a system cannot rearrange itself to this configuration and so is not in the relative equilibrium, it will still rotationally fission but at a higher spin rate. The system will then undergo the same dynamics demonstrated below but start with a higher energy increasing the probabilities of disrupting the system and a re-impact event between the two components. An impact event will dissipate energy but conserve angular momentum, so there is no stable single-body configuration for the system, and thus material must immediately lift off the primary and return to orbit~\citep{Scheeres2009a}. The specific details of the impact and ejecta will determine whether the system evolves as a high or low mass ratio system. For the sake of simplicity, the model will assume the minimum energy configuration with the knowledge that the results have a systematic uncertainty due to this initial condition.

\subsection{Rotational Fission Model}
\label{sec:rotationalfissionmodel}

[Figure 2]

The dynamical simulation begins with a tri-axial asteroid made of two components inspired by objects such as Toutatis or Itokawa. We model ``rubble pile'' asteroids as having an inherent component mass ratio dividing all of the ``rubble'' into hierarchical groups determined by the largest distance between mass centers as shown in the upper left panel of Figure 2. The mass ratio between the components and the shape ratio of each component are the three initial free parameters. As the asteroid's rotational rate increases due to the YORP effect, the long axes of each ``rubble pile'' component ellipsoid will align for rapid rotation rates. This configuration is the only stable relative equilibrium figure for the body while still resting on each other~\citep{Scheeres2007b}. As the YORP effect continues to torque the body, the two components will enter into orbit about each other.

The YORP effect is responsible for spinning the initial body up to the required rotation rate for fission. This is the only time a non-gravitational process (Yarkovsky, YORP, or BYORP) is required for constructing synchronous binary systems. The components will fission at a specific spin rate of the primary body given the internal component mass distribution~\citep{Scheeres2009a}. All bodies in these simulations have a density of 2 g/cc and so will fission at rotational periods greater than 2.33 hours dependant only on the mass and shape ratios of the two model components.

\subsection{Dynamical Model}
\label{sub:DynamicalModel}

Post-rotationally fissioned systems were studied by directly integrating the Lagrangian dynamics with an implicit 12th order Runge-Kutta scheme for two bodies (two tri-axial ellipsoids) or three bodies (one tri-axial ellipsoid and two spheres) as described in~\ref{sec:derivationtwobody} and~\ref{sec:derivationthreebody}, respectively. Non-spherical gravitational potentials ensure that the model will capture the important effect of spin-orbit coupling. The model incorporates secondary fission (selecting secondary component mass ratios from a flat distribution between $0.01$ to $0.99$) and impacts (inelastic collisions with total angular momentum and mass conserved) as described in~\ref{sec:derivationsecondaryfission} and~\ref{sec:impactmassredistribution}, respectively. The dynamics also include the torques from mutual body tides, which dissipate energy. The rate of energy dissipation is dependent on the difference between the spin rate of the body and the orbital rate and inversely dependent on the distance between the bodies to the sixth power~\citep{Murray1999}. The specifics of the tidal theory is given in~\ref{sec:derivationtidaltheory}. The effect of solar gravitational perturbations for an orbit about the Sun at 1 AU is also included on the system. In each integrator, total system energy is conserved to greater than 1 part in $10^8$ when energy changing effects such as mutual body tides and solar gravitational perturbations are neglected, and angular momentum is conserved to greater than $1$ part in $10^8$ when angular momentum changing effects such as solar gravitational perturbations are neglected.

Non-gravitational forces (YORP and BYORP effects) were not included in the post-rotational fission dynamical model since the gravitational timescales are much shorter than the radiative timescales. The model assumes planar motion with the intention of implementing 3-D motion and associated non-principal axis rotation in the future, with the expectation that it will increase energy dissipation, lengthen the timescale for ejection of the secondary, and thus increase the binary formation efficiency; more detailed modeling of impact and fission processes may change the efficiencies associated with the evolutionary sequence but should not change the possible outcomes. The dynamics will scale to any realistic size scale for NEAs, although the effects of cohesive attraction are not modeled and may be important on the smallest size scales $\lesssim 100$ m~\citep{Scheeres2010} and the timescale for YORP effect induced rotational fission is too long for the largest size scales $\gtrsim 10$ km.  

\section{Results}
\label{sec:results}

\subsection{Chaotic Binary}
\label{sec:chaoticbinary}

The dynamics demonstrated immediately after the initial rotational fission are chaotic. The coupling of the spin and orbit states from the tri-axial gravitational potential is responsible for significant variations of behavior in the system. This coupling can transfer large amounts of energy and angular momentum across the system leading to rapid changes in the spin rates and orbital revolution rates. These changes can repeatedly switch the tidal bulge on each member from leading to lagging and vice versa. Immediate tidal energy dissipation via mutual body tides helps prevent re-impact. 

Studying two cases in detail helps illustrate this chaotic evolution. These systems approximate the well-known systems: (66391) 1999 KW$_4$, an asynchronous binary, and 25143 Itokawa, a contact binary.~\citet{Ostro2006} reports the size of the primary of 1999 KW$_4$ as close to a tri-axial ellipsoid with semi-axes $708.5\times680.5\times591.5$ m, and the secondary as a tri-axial ellipsoid with semi-axes $297.5\times225\times171.5$ m. Assuming constant density, the mass ratio of the binary system is close to $0.04$.~\citet{Demura2006} reports that size of the ``body'' (primary) of Itokawa approximates a tri-axial ellipsoid with semi-axes $245\times130\times130$ m, and the ``head'' (secondary) as a tri-axial ellipsoid with semi-axes $115\times90\times90$ m. Assuming constant density, the mass ratio of these components is close to $0.3$.

For these simulations the bodies are placed into their relative equilibrium state, which is the only stable configuration of the bodies when the rotation rate of the system is just below rotational fission of the two components~\citep{Scheeres2009b}. In this configuration, the primary and secondary are in contact and aligned along their longest axes. The dynamical simulation begins when the rotation rate of the pre-fissioned body reaches the critical disruption spin limit for the two, internal components. The dynamics modeled include spin-orbit coupling and mutual body tides as described in Section~\ref{sub:DynamicalModel}, but do not include solar gravitational perturbations, so angular momentum is conserved.

[Figure 3]

Exploring the evolution of the semi-major axis $a$ and eccentricity $e$ for 1999 KW$_4$ and Itokawa over the first 60 days after rotational fission in Fig. 3, the nature of the chaotic evolution is evident. Variations in the semi-major axis track transfers of angular momentum from the orbital state to the spin state and back. The spin of each body is coupled to the orbital motion through the non-Keplerian gravitational potential, and these transfers of angular momentum are apparent in the changing rotational periods of the primary and secondary of each system as shown in Fig. 4. The spin rates of both bodies change dramatically, however as the mass ratio decreases the changes in the spin rate of the secondary become more dramatic. The secondary of 1999 KW$_4$ has intervals of almost no rotation, but also intervals when its rotation rate exceeds and resides near the critical disruption limit for a sphere. The critical disruption limit for a sphere is defined as the rotation rate necessary to lift a massless test particle off the surface, if the secondary had a ``rubble pile'' internal structure, then it would disrupt at a slower rotation rate. Rotational fission of the secondary due to torques from spin-orbit coupling is called secondary fission and occurs in many simulated systems. This process is discussed further later. If the secondary of 1999 KW$_4$ is not allowed to fission and the system continues to dynamically evolve, then the system will disrupt after $\sim1600$ days. The Itokawa system will never disrupt since the system has a negative free energy and so is always bound.

[Figure 4]

This chaotic evolution causes the rate of tidal energy dissipation to change radically between intervals of strong and weak or even non-existant tidal dissipation, which depends on the separation distance to the sixth power and the relative spin rates of the bodies to the orbital rate. While the motions of an individual system are chaotic, the systems as a whole do appear to have a general set of dynamics and show trends with mass ratio. The secondary vacillates between libration and circulation more often than the primary which usually displays just circulation. The system evolves rapidly into an eccentric mutual orbit with a quickly changing longitude of pericenter. Secondaries of lower mass ratio systems exhibit stronger instances of these behaviors. This is expected because the lower the mass ratio, the greater the initial spin rate for rotational fission in the parent body. Lower mass ratio systems will thus have more energy to transfer into their orbits after fission. 

[Figure 5]

\subsection{Two Regimes}
\label{sec:tworegimes}

The initial spin rate for rotational fission in the parent body quickly divides the dynamics into two regimes: negative and positive free energy. The initial rotation rate necessary to fission the components depends on the mass ratio between the two components~\citep{Scheeres2009b}. For two spherical components the division between the two regimes occurs at a mass ratio of $\approx 0.2$. High mass ratio systems (mass ratio $>0.2$) have a negative free energy and are bound for all time under internal gravitational perturbations. Low mass ratio systems (mass ratio $<0.2$) have a positive free energy and may escape if the excess energy is not dissipated. For tri-axial ellipsoids the regime boundary does not have a specific value since there is a dependence on the shape of each body, however the results indicate that the regime boundary is real and can be approximated by a mass ratio of $0.2$. Increasing the shape ratio (elongating the objects) decreases the mass ratio of a zero free energy system, but not by much~\citep{Scheeres2009b}.

Fig. 5 shows the time averaged separation distances of $150$ systems with mass ratios from $0.05$ to $0.99$ after $100$ years. The Hill radius ($80.5$ primary radii at $1$ AU) determines boundedness for these systems. The regime change between bound (high mass ratio) and unbound (low mass ratio) is dramatic and has consequences for the subsequent evolutionary path of the systems.~\citet{Pravec2010} has directly observed this mass ratio spin limit in the asteroid pairs population, which is discussed further in Section~\ref{sec:asteroidpairs}. 

Both high and low mass ratio regimes have chaotic early dynamics. These dynamics increase the eccentricity and thus the energy dissipation from mutual body tides. High mass ratio systems will evolve to an orbital equilibrium state, but low mass ratio systems will disrupt before tidal dissipation can reduce the free energy to a negative value unless they undergo secondary fission. The outcomes of these processes on each of the regimes are detailed below. 

\subsection{High Mass Ratio Regime}
\label{sec:highmassratioregime}

[Figure 6]

High mass ratio systems are defined as those systems that have negative free energy and do not experience secondary fission. The upper branch of Fig. 1 shows the evolutionary path of high mass ratio systems. The dynamics of these systems are chaotic, but since the bodies are more equal in size, the exchanges of angular momentum and energy through spin-orbit coupling are less severe. This inhibits secondary fission and reduces the eccentricities these systems experience as they evolve. The tidal energy dissipation rate is inversely related to the separation distance to the sixth power, and so higher mass ratio systems experience faster rates of tidal energy dissipation, since the average separation distance decreases with higher mass ratios, as shown in Fig. 5.

Tidal dissipation damps systems in the high mass ratio regime so that both the primary and secondary of such systems are trapped in libration states. The libration angle is eventually damped to zero, first in the secondary then the primary, and the bodies become doubly synchronous. Since high mass ratio systems have similarly sized components, the tidal timescale is similar for each member and systems evolve into doubly synchronous binaries. Tidal timescales are a direct function of mass ratio with equal mass members taking $\sim 5 \times 10^3$ years, $0.6$ mass ratio systems taking $\sim 10^4$ years, and $0.2$ mass ratio systems taking $\sim 2 \times 10^6$ years to reach the doubly synchronous state (see~\ref{sec:modelingtidaltimescales} for a description of the assumptions behind these timescales). Fig. 6 shows these timescales as a function of mass ratio, along with a fitted power law showing a clear trend as a function of the system's mass ratio. Both members are tidally locked in the doubly synchronous state, and these asteroids are observed as the Hermes-class. 

This numerical tidal dissipation timescale for high mass ratio systems can be compared to the tidal timescales derived analytically by Goldreich and Sari (2009). The corresponding analytic tidal dissipation timescales for equal mass members is $\sim 3 \times 10^4$ years, for $0.6$ mass ratio systems is $\sim 4 \times 10^4$ years, and for $0.2$ mass ratio systems is $\sim 4 \times 10^6$ years to reach the doubly synchronous state. The analytic theory is within an order of magnitude of the numerical results but consistently overestimates the time necessary to de-spin these systems especially at higher mass ratios. Fundamentally, the analytic theory assumes a quasi-steady state evolution, but this is not how these systems initially evolve. All of these systems engage in a period of chaotic evolution that can increase the spin rates of the bodies relative to the orbit, and since the tidal energy dissipation rate is linearly related to the difference between the spin rate of the body and orbital rate, the energy dissipation is faster than that predicted by the analytic theory.

Once in the doubly synchronous state, the system will contract or expand due to the BYORP effect creating contact binaries or asteroid pairs~\citep{Cuk2007, McMahon2010}. The BYORP effect is the summation of radiative effects on synchronous secondaries. It can shrink or expand the semi-major axis. If the semi-major axis expands, the asteroid system will eventually disrupt when the separation distance equals the Hill radius. If the semi-major axis shrinks, the two components will at first remain in the doubly synchronous state since this is also the stable relative equilibrium state until the separation distance reaches a lower limit and the relative equilibrium state becomes unstable~\citep{Scheeres2009b}. Simulations show the impacts occur very soon ($<100$ days) after reaching the stability limit, and the perpendicular and tangential impact velocities for $1$ km bodies are $< 50$ mm/s, modest enough to be capable of creating contact binaries.  These impact velocities are gentle enough that they would not disrupt the figure of the bodies creating the contact binaries. Thus, high mass ratio evolution is responsible for creating doubly synchronous binaries, which can evolve into contact binaries or asteroid pairs.

The end products of this sequence are single ``rubble pile'' asteroids, so this is a lifecycle. We propose a possible contact binary loop that high mass ratio systems could get stuck in, whereby the components of a contact binary repeatedly fission and re-impact. Each component maintains its relative orientation to the other, so that the YORP effect, BYORP effect, and mutual body tides all act similarly each time the system goes through the cycle. The estimated timescales of the tidal process and the BYORP process are roughly an order of magnitude shorter than the estimated timescale of the YORP process \citep{Rossi2009, McMahon2010b}. This would explain why contact binaries are so prevalent compared to doubly synchronous systems. Contact binaries appear to be $9\%$ of the NEA population, and very roughly this theory would predict that the doubly synchronous population would be $\sim5$ times smaller due to timescales. Also, this theory would predict that the ratio of timescales would reflect the ratio of contact binaries to doubly synchronous systems with a caveat regarding asteroid pair production from doubly synchronous systems. 

\subsection{Low Mass Ratio Regime}
\label{sec:lowmassratio}

Low mass ratio systems have positive free energy or undergo secondary fission. These systems typically have mass ratios $<0.2$. Coupling between the spin and orbit states drives a large spin increase in the secondary and an increase in the eccentricity of the system. These systems chaotically explore their phase space until an escape trajectory is discovered or secondary spin fission occurs, which will be further defined in Section~\ref{sec:secondaryspinfission}. 

If low mass ratio systems are evolved after a rotational fission event and they are not allowed to secondary spin fission, then almost all systems will disrupt. $450$ low mass ratio systems were simulated starting from rotational fission and evolved considering tri-axial gravitational potentials, mutual body tides, and solar gravitational perturbations starting from the relative resting equilibrium as described in Section~\ref{sec:methods}. The systems were evolved until they disrupted, which is when the separation distance equals the Hill radius. The Hill radius is taken to be 80.5 primary radii, which is correct for a system is in a circular heliocentric orbit at 1 AU, but mutual orbits that grow to this separation distance typically reach much larger separation distances if the system is allowed to continue to evolve.

[Figure 7]

The time to system disruption is shown in Fig. 7, firstly as a function of mass ratio and secondly as a function of primary shape ratio, defined as the shortest semi-axis divided by the longest semi-axis of the primary in the plane of motion. Similarly, a secondary shape ratio can be defined. The initial component mass ratios, which determined the binary mass ratios, were chosen from a flat distribution ranging from $0.001$ to $0.2$. Although secondary fission was not allowed to occur in these simulations, each binary was still defined using the hierarchical ``rubble pile'' internal structure as shown in Fig. 2. Therefore, the internal component mass ratio of each binary component is chosen from a flat distribution between $0.001$ and $q/(1-q)$, where $q$ is the mass ratio of the previous fission, this requirement is derived in~\ref{sec:massratiosecondaryfission}. The internal component mass ratio also determines the shape ratio of the component, since each component is dynamically modeled as a tri-axial ellipsoid with the same moments of inertia as the hierarchical ``rubble pile'' internal structure model. Thus the primary shape ratios are distributed between $0.581$ and $0.997$.

$2 \pm 1\%$\footnote{All uncertainties from the model given as $\# \pm \#$ describe the most likely actual proportion and an estimate of the sampling error given as a $90\%$ confidence interval. These likelihoods and confidence intervals are calculated using the Wilson Score Confidence Interval, which best approximates a binomial distribution especially at extreme probabilities--small number of successes compared to number of trials~\citep{Agresti1998}.} of systems integrated, $7$ out of $450$, do not disrupt after $1000$ years of integration. The simulation was ended after $1000$ years, because the timescale of the BYORP effect becomes comparable and this effect was not included in the simulation. The $7$ binary systems that did not disrupt evolved very differently than the other $443$ systems; the secondary rotation and orbital periods remained very close to one another, slowly growing as mutual body tides dissipated energy. The primary rotation period is slightly smaller than the other periods as the system evolves. 

These binaries evolved differently than the other systems because their primaries were significantly more spherical as shown in Fig. 7. All surviving binaries had primary shape ratios greater than 0.98 and all disrupted systems have primary shape ratios below 0.98. The mass ratios and secondary shape ratios do not effect the outcome of the evolution of the system. When $50$ more systems were evolved with primary shape ratios restricted to lie between $0.9$ and $1$ and the mass and secondary shape ratios varied over the same range as before, these conclusions are modified slightly. The boundary between disruption and stability is not as sharp, one system with a primary shape ratio below 0.98 does not disrupt (it had a primary shape ratio of 0.978), and two systems with primary shape ratios greater than 0.98 do disrupt (they had primary shape ratios of 0.981). No observed binaries have primaries with primary shape ratios of 0.98 or greater. The largest known primary shape ratio is 0.96~\citep{Pravec2007}, as shown with other observed binaries in Fig. 13. Therefore, the $2\%$ likelihood of creating a stable binary that evolves along the unstable relative equilibrium is strongly dependent on the assumption of a flat distribution in primary shape ratio, and potentially no primaries after a rotational fission event may ever be that close to spherical.

These systems are evolving outward along a relative equilibria as theoretically shown in ~\citet{Scheeres2007b}, where the case of a small ellipsoid rotationally fissioning from a large sphere is explored. In~\citet{Scheeres2009b}, it was shown that this relative equilibrium would always been unstable for two ellipsoids, however if the primary shape ratio is nearly one, then the growth of this instability may be slow compared to the tidal dissipation. In this case, the system may evolve outward along the relative equilibrium without chaotically fully exploring its phase space, instead it may only exchange limited angular momentum and energy between the orbit and rotation states causing libration and circulation in the primary, libration with very rare circulation in the secondary, and small changes in eccentricity and semi-major axis. As the system evolves, tidal energy dissipation will slowly grow the pericenter, and since the higher order gravitational potential terms have a  $1/r^3$ functional dependence, the effects of the small non-sphericity of the primary will diminish even more. The secondary rarely circulates and is often librating with a very small angle, so the BYORP effect would significantly effect the evolution of the system in potentially only a few thousand years~\citep{Cuk2007,McMahon2010b}. The properties of these binaries that survive without disruption are shown in Fig. 13 and more discussion of  their continued evolution is further discussed in Section \ref{sec:stablebinaries}. 

For those systems that do disrupt, the median time to disruption for all systems is $32_{18}^{72}$ days\footnote{All statistics reported from the model are given as $\#_\#^\#$ and describe quantile statistics that enclose~$50\%$ of the data, since the underlying distribution is unknown. In the normal script is the median value, and then in the subscript is the 25th percentile and in the superscript is the 75th percentile.}. An exponential decay can be fit to the data $N(>t) = 443 e^{-t/\tau}$, where t is the time after rotational fission, $N(>t)$ is the number of asteroid systems remaining after time $t$, and $\tau = 92.0\pm1.8$ days is the exponential decay timescale. The adjusted $R^2$ value of the fit is $0.996$. The half-life to disruption for low mass ratio systems is then $\tau_{1/2} = 63.8\pm1.2$ days.

As shown in Fig. 7 and also shown in a simple binning of the data as done in Table 2, there is a trend in the disruption time with the mass ratio. The lower the mass ratio the shorter the median time to disruption. This is a direct result of the added energy necessary to initially fission lower mass ratio systems.  Higher mass ratio systems experience lower average eccentricity, explore their orbital phase space more slowly, and thus can find disruption orbits on much longer timescales. A stronger trend than the dependence on mass ratio is the dependance on primary shape ratio. The lower the primary shape ratio the shorter the median time to disruption, despite the energy necessary for rotationally fissioning a system decreasing with a smaller primary shape ratio~\cite{Scheeres2007b}. This relatively small effect is strongly counteracted by the increase in the size of the second order terms in the gravitational potential, which increase the coupling of the spin and orbit states. The spin-orbit coupling through these non-spherical gravitational terms is how energy is transferred into the orbit from the rotation states eventually disrupting the system. A useful analogy is the time it takes Theseus to escape the Cretan Labyrinth; the number of exits from the maze and the speed at which Theseus explores different passages increases with the decreasing mass ratio and decreasing primary shape ratio of the chaotic asteroid binary. These trends are nonlinear and appear logarithmic. The disruption time appears to a approach a constant value as the mass ratio and primary shape ratio approach zero. There appears to be no or a very weak trend in the disruption time with secondary shape ratio.

The disruption timescales for rotationally fissioned systems are very short compared to the YORP timescales for fissioning the primary again before the system disrupts. This is true even for systems with primaries that are more rotationally symmetric than any of the observed primaries of binary asteroid systems. Stabilization of the secondary via collision with more material fissioned from the primary would require extremely (and unobserved) large YORP accelerations and hence very short YORP timescales. Something else must happen to the system before disruption, in order to form synchronous binaries. We hypothesize from our numerical modeling that this process is spin fission of the secondary.

[Table 2]

\subsection{Secondary Spin Fission}
\label{sec:secondaryspinfission}

Rotational fission rests on the premise that asteroids are ``rubble piles,'' and so this naturally leads to the assumption that the primary and secondary members of the chaotic binary formed from rotational fission are also ``rubble piles.'' During the evolution of the two-body system, spin-orbit coupling can increase the spin rate of the secondary such that it undergoes rotational fission of its ``rubble pile'' structure. Both asteroids undergo surface fission at similar rotation rates, however because of the large mass difference between the bodies they disrupt at very different rotational kinetic energies. It takes much less energy transferred via spin-orbit coupling to the secondary to fission that body. 

The most conservative scenario for secondary spin fission is surface fission--the condition for a massless test particle resting on the surface of the secondary to become unbound. A real massive component would become unbound at some lesser condition as described in~\ref{sec:derivationsecondaryfission}. The full two-body integrator checked the surfaces of each asteroid as it evolved for this condition at every time step, and then implements Brent's method (a bracketed root finding method) to determine the time and state of the system when the condition is first satisfied.

During the evolution of the $443$ low mass ratio systems simulated above, $178$ undergo surface fission of the secondary before orbital disruption. That is $40\pm 4\%$ of the modeled systems (uncertainties attained using the Wilson Score Confidence Interval). Secondary surface fission is a conservative limit that corresponds to the spin rate necessary to place a massless test particle on the surface of the original body into orbit. If these secondaries have ``rubble pile'' geophysics then they would secondary fission at lower spin rates~\cite{Scheeres2007b}. This hypothesis is pursued later in the numerical simulations discussed in Section~\ref{sec:chaotictertiaries}.

[Figure 8]

In Fig. 8, those systems are shown as crosses at the time of secondary fission, while those that did not secondary fission are shown as dots at the time of disruption. For those systems that underwent secondary fission, the median time to surface fission was $51_{27}^{128}$ hours. Those systems that take the longest to disrupt are also the most likely to go through secondary fission.  Spin-orbit coupling transfers free energy throughout the system temporarily storing it in different reservoirs such as the spin states of the bodies at different times. If too much energy is stored in certain kinetic energy reservoirs, the system can be irreversibly changed. These two reservoirs are: the spin energy of the secondary and the relative translational energy of the bodies. If too much energy is stored in the translational energy the system will disrupt, and if too much energy is stored in the spin of the secondary than the secondary will fission. Fission of the primary is theoretically possible, however it was never observed in the numerical experiments. While the rotation rate needed to surface fission the primary is the same as the secondary--the surface fission rate only depends on density~\citep{Scheeres2007b}, the rotational kinetic energy necessary to achieve that rotation rate is much higher, and a transfer of this much energy into the rotation state of the primary does not occur.

Fig. 8 also shows how the likelihood of secondary fission depends on the three parameters in the simulation. The fraction and errors shown in the plots on the right hand side of Fig. 8 are given for bins of width $0.025$ for mass ratio and $0.05$ for each of the shape ratios and were calculated using the Wilson Score Confidence Interval. The fission fraction decreases with increasing mass ratio, which follows naturally from the decreasing amount of energy necessary to initially rotationally fission the original body. If there is less energy to transfer via spin-orbit coupling through the binary system, then there will be less energy to momentarily store in the rotation reservoir of the secondary. The fission fraction decreases with larger secondary shape ratio. The smaller second order gravitational potential terms of the secondary lower the coupling of the spin state to the orbit state decreasing the ability of energy to be transferred into the rotation rate of the secondary. The fission fraction decreases with smaller primary shape ratio for a very related reason. The increased spin-orbit coupling of the primary to the orbit increases the energy transferred into the orbit and increases the semi-major axis so that the secondary is prevented from being rotationally accelerated.

\subsection{Chaotic Ternaries}
\label{sec:chaotictertiaries}

Secondary spin fission drastically alters the evolution of the system. After secondary fission, the asteroid system is now a chaotic ternary. These systems could stabilize via tidal dissipation into the observed ternary asteroid systems, but more likely one of the secondaries will either exit the system: further stabilizing the orbit of the secondary through removal of energy and angular momentum, or impact the primary: increasing its spin rate and potentially creating an equatorial bulge. This process provides a route to the creation of synchronous binaries.

Secondary fission often occurs when the orbit of the secondary is at pericenter and the location of the fission is on the interior (primary facing) side of the secondary. The fissioned material will be at apoapse of a new orbit with periapse close to or inside the primary and so this material will quickly impact the primary. These impacts have speeds $< 1$ m/s, and so will not disrupt the primary, but may re-organize it's shape. This mechanism may be responsible for forming the observed equatorial bulges seen in the near-Earth asteroid population, specifically primaries of synchronous binaries and fast-spinning, single asteroids. Impacts are modeled to conserve angular momentum and mass, and the collision is treated as inelastic as described in~\ref{sec:impactmassredistribution}. 

The remainder of the secondary still in orbit is now at periapse of a new larger orbit that is more stable. The secondary fission may repeat many times during the evolution of a system. The model also incorporates tidal effects including solar gravitational perturbations and mutual body tides, which work over time to circularize and synchronize the secondary to create the observed synchronous binaries. The gravitational effects of the Sun on the mutual orbit provide important stability to the system during the transition period between chaotic evolution and quasi-steady state evolution dominated by mutual tidal dissipation.

This process was modeled using the full two-body integrator, as above and initial rotational fission component mass ratios were drawn from a flat distribution between $0.01$ and $0.2$ to capture the complete low mass ratio regime. After the initial rotational fission the secondary is treated as ``rubble pile'' itself with initial component mass ratios chosen from a flat distribution between $0.01$ and $0.99$\footnote{A flat distribution across all possible values is the simplest assumption. In Section~\ref{sec:secondaryspinfission}, we chose the more conservative assumption that results in higher secondary fission ratios given by~\ref{sec:massratiosecondaryfission}.}. The asteroid is treated as a hierarchical structure as depicted in Fig. 2, only the next . Until the secondary fission condition is met, each component of the binary is treated as a coherent dynamical body. Once the secondary fission condition is met, then the system becomes a chaotic ternary. After one of the chaotic ternary members is ejected from the system or impacts another member, then the smaller of the two remaining members is once again treated as a ``rubble pile'' with a component mass ratio chosen from a flat distribution with an appropriate upper limit and subject to the secondary fission condition. 

The exact condition for secondary fission is described for the case of a primary tri-axial ellipsoid and a secondary composed of two spheres in~\ref{sec:derivationsecondaryfission}. The motion of the chaotic ternary, which is made of one tri-axial ellipsoid and two spheres is determined by the three-body integrator described in~\ref{sec:derivationthreebody}. This integrator includes the effect of solar gravitational perturbations for a circular orbit about the Sun at 1 AU and mutual body tides between the primary and each of the satellites. The three-body system is highly chaotic and rapidly explores its phase space. There are four possible outcomes to this state: stable ternary system, collision between the spherical bodies, impact of one spherical member onto the tri-axial primary, and escape of one of the spherical bodies.

 $526$ systems were evolved for $1000$ years unless the system disrupts or the components impact each other. The model defines stable ternary systems as those that last until the end of the simulation, $1000$ years ($\sim 10^4$ disruption timescales). Since two of the three members are perfect spheres not all of the dynamics are captured by the three-body integrator and these systems are stable in only a limited sense. The three remaining paths are diagrammed in Fig. 10. The first path is ejection of one of the satellites. This results in a binary system with a spherical secondary, but to capture the spin-orbit coupling the secondary is then given some small prolate-ness and a new internal ``rubble pile'' structure (shape and mass ratios are randomly drawn from the same initial, flat distributions). The dynamics are then returned to the two-body integrator and the system continues to be evolved. The second path from a chaotic ternary to a chaotic binary is impact of one component of the secondary onto the primary. These impacts occurs at low speeds $<1$ m/s. When this occurs the mass and angular momentum are conserved and the collision is treated as inelastic as described in~\ref{sec:impactmassredistribution}. The third path is impact of the two satellites with each other and these impacts are treated in the same way. From simulation, the impact velocities of both of these impacts suggest that single re-shaped asteroids are a more likely outcome rather than fragmentation. These velocities are described later in this work. 

After a binary system has gone through secondary fission, it will have lost energy which raises the semi-major axis and thereby increasing the periapsis distance. The higher-order gravitational effects responsible for spin-orbit coupling are strongest at periapse, and so these systems will not be as affected by spin-orbit coupling. However, the system will often still be eccentric and solar gravitational perturbations are important for stabilizing and destabilizing these systems. Solar tides will change the energy and angular momentum of the system and when they stabilize the orbit they do so by expanding the pericenter and lowering the apocenter, which keeps the system from impact and disruption.

[Figure 9]

After $1000$ years of evolution, $8 \pm 2\%$ of low mass ratio systems are stable binaries, $67 \pm 3\%$ of simulated systems disrupt and become asteroid pairs, and $25 \pm 3\%$ of the simulations end with the secondary impacting the primary at modest speeds creating re-shaped asteroids. While ternaries exist in some simulations for a number of years, none of the systems remain as stable ternaries; this sets an upper limit on the likelihood of stable ternary formation at $0.3 \pm 0.3\%$ for our simplified model. These intervals capture the statistical or random errors but do not include systematic effects from broad assumptions such as the internal components size distribution (we assumed a flat distribution). Other systematics such as the assumption of planarity and a full body model for the three body system will be developed in the future. There is also another route to stable ternaries, where the primary of a stable binary goes through YORP-induced fission and the system may evolve such that this new ternary system does not disrupt. These systems were not modeled here.  

[Figure 10]

Secondary fission is a dominant process--$64 \pm 3\%$ of low mass ratio systems go through it at least once. Secondary fission creates a chaotic ternary that evolves back into a binary via one of the three routes. One of the secondaries impacts the primary $51 \pm 3\%$ of the time. Otherwise, one of the secondaries is ejected from the system $28 \pm 2\%$ of the time, or the secondaries collide in orbit about the primary $21 \pm 2\%$ of the time. All of the these chaotic ternary processes dissipate energy and produce more stable binaries. 

Collisions between the secondaries of a ternary system occur in $17 \pm 3\%$ of low mass ratio systems at least once. Impacts also occur on $83 \pm 3\%$ of low mass ratio primaries via either a ternary component impact after a secondary fission or collision with the secondary in a binary. Each of the three collision processes has a unique velocity structure shown in the top plot of Fig. 11. The median velocity, as well as first and third quartiles are listed in Table 3 for all three processes. Each type of impact has a unique tangential velocity, but similar perpendicular velocities.

[Table 3]

The bottom two plots of Fig. 11 compare the impact velocities that occur in the model to catastrophic disruption limits derived elsewhere. The triangles and dots are piled up in both plots. The middle plot shows the total velocity of the impactor relative to the target and compares this value to the catastrophic disruption velocity of~\citet{Stewart2009}. The upper line is the critical velocity for projectiles much smaller than the target, which corresponds with impacts onto the primary of either the binary (triangle) and ternary (dot) systems. The lower line is the critical velocity for nearly equal size projectile and target. The collisions between secondaries of ternary systems do not fall entirely into either domain; they have mass ratios between $0.01$ and $0.99$. From this analysis a minority of collisions may undergo catastrophic disruption, however analysis of the specific energy of these collisions leads to the opposite conclusion. The bottom plot of Fig. 11 shows the specific energy of each impact and compares this to the catastrophic disruption limits derived by others. The two curved and dotted lines indicate models that include both internal strength and self-gravity, the two dashed lines only include self-gravity, and the dot-dashed line represent the self-potential energy of the target for reference. With the exception of a single impact, these collisions do not catastrophically disrupt the target body, but will re-arrange material on the surface. For impacts on the primary, material will preferentially impact the equator and we hypothesize that it creates the commonly observed equatorial bulge and should fill in gravitational potential lows circularizing the body and further stabilizing the dynamics.

[Figure 11]

\subsection{Stable Binaries}
\label{sec:stablebinaries}

[Figure 12]

There are $41$ systems out of $526$ forming stable binaries at the end of the post-rotational fission dynamical simulation. They have a median mass ratio of $0.003_{0.001}^{0.008}$ and a median semi-major axis of $3.3_{2.6}^{6.0}$ primary radii. The distribution of semi-major axes is shown in the top panel of Fig. 13. There is an excess of large semi-major axes compared to the observed systems, however the BYORP effect has not been taken into account, and it will move the semi-major axis of nominally half of these systems inward. The output binary systems also have eccentricities with a median of $0.32_{0.15}^{0.45}$ that will also be tidally damped over time.

[Figure 13]

The median primary rotation period is $3.9_{3.5}^{4.6}$ hours and are shown in the middle plot of Fig. 13. Compared to the observed systems, the modeled primaries are spinning slowly, however the YORP effect will continue to spin up the primary and while this effect operates on a timescale longer than the simulation, the YORP timescale is short compared to the lifetime of the asteroid system.

The secondaries have fast spin periods, spun up via spin-orbit coupling; the median secondary spin period is $2.8_{2.1}^{3.7}$ hours. The distribution is shown in the lowest panel of Fig. 13. $29 \pm 5\%$ of secondaries are retrograde after the chaotic phase. The critical disruption limit only includes mass ratios $>0.01$, which corresponds to secondaries $\gtrsim 100$ m in radius, which are either coherent bodies with internal strength or ``rubble piles'' bound by cohesive forces. The stable binaries output by the simulation spin too fast compared to the observed synchronous binaries, however tidal dissipation will synchronize these systems.

[Figure 14]

The timescale for a low mass ratio secondary of a binary asteroid system to evolve to the synchronous state, $\tau_\text{tidal}$, is estimated by dividing the spin rate, $\omega$, by the magnitude of the tidal acceleration of the secondary, $|\dot{\omega}_\text{tidal}|$.

\begin{equation}
\tau_\text{tidal} = \frac{\omega}{|\dot{\omega}_\text{tidal}|} = \frac{Q}{5 k} \left( \frac{\omega}{G \pi \rho} \right) a^6
\end{equation}

\noindent where $k = 10^{-5}$ is the tidal Love number, $Q = 100$ is the quality factor, $G$ is the gravitational constant, $\rho = 2$ g cm$^{-3}$ is the asteroid average density, $a$ is the binary system semi-major axis in units of primary radii \citep{Goldreich1963, Goldreich1966}. Assuming rubble pile geophysics, eccentricity will damp for all mass ratios \citep{Goldreich2009}. The secondaries, due to their much smaller dimensions and slower relative spin rate, tidally damp before the primary and create synchronous binaries  \citep{Goldreich2009}. Assuming a primary radius of 1 km, the median estimate of the tidal spin-down timescale for the created synchronous binaries is $1.6 \times 10^5 \ _{3.0 \times 10^4}^{1.7 \times 10^6}$ years, as shown in Fig. 14.  The analytic theory of \citet{Goldreich2009} predicts a median timescale for the observed synchronous asteroid population within an order of magnitude of the simulated systems, $1.7 \times 10^5 \ _{1.1 \times 10^5}^{2.6 \times 10^5}$ years, but a very different dispersion. This dispersion is the result of a few simulated systems having very large and very small secondary spin periods. Those systems with very small secondary spin periods may not synchronize within their lifetime in the NEA population and may be observed as asynchronous, high-$e$ binaries, however these systems seem to only exist for very low mass ratio. 

The YORP effect will also evolve the spin state of the primary and secondary, however the YORP effect will be stronger on the smaller secondary and potentially match the timescale for tidal synchronization. The evolution of the secondary may follow three paths depending on the relative directions and strengths of the YORP effect and tides. Firstly, tides and YORP act in the same direction lowering the timescale to synchronization. Once the system is synchronized, the YORP effect will not be able to remove it from that state, because the YORP torques will be much less than the gravity gradient at the separation distances of the observed and simulated systems. The YORP effect may provide a small source of angular momentum through the secondary to the system creating a small leading offset in the orientation of the secondary and very slowly evolving the orbit. The other two ways in which the secondary may evolve occur when the YORP effect and tides. If tides dominate in strength, then the system will synchronize and the YORP effect will act as a small sink of angular momentum causing the system separation to shrink and a small trailing offset in the orientation of the secondary. If the YORP effect dominates, then synchronization may never occur and the BYORP effect will never evolve the system. Tides will continue to expand the system, but the YORP effect will keep the secondaries rapidly rotating. This last path may be responsible for the as asynchronous, high-$e$ binary systems as well, but without the mass ratio dependance apparent in the previously mentioned source of high-$e$ binaries.

The chaotic evolution of low mass ratio systems including the effects of secondary fission and the ensuing consequences of impacts and escapes drastically changes the initial spin rate of the secondary for the classical tidal theory, which assumes that the secondary starts at the spin fission limit. Chaotic evolution of the secondary appears as a random walk in spin rate. If the secondary walks to higher spin rates it will eventually spin fission, and the random walk will be reset for each of the secondaries. If it walks to slower spin rates, then when the system evolves into classical tidal evolution, the system will de-spin on a shorter timescale. This would be true of all stable binaries output by the dynamical model, however some systems form stable binaries with large semi-major axes. Classical tidal dissipation is inversely related to the separation distance to the sixth power, and so these systems may only be a factor of a few larger in semi-major axis, but that translates into a difference of over two orders of magnitude in tidal dissipation rates. Those systems that take longer than or similarly to the lifetime of an NEA system ($\sim 10^7$ years) to synchronize do not become synchronous binaries, instead they become the rarer high-$e$ binaries. 

Once a system is synchronized, the BYORP effect can contract or expand their orbit. Synchronous binaries disrupt once the orbit has expanded to the Hill radius creating asteroid pairs. The BYORP effect can also contract the orbit to the stability limit  leading to the secondary impacting the primary, re-shaping the body due to the primary's rapid rotation rate and creating re-shaped asteroids~\citep{McMahon2010b}. Thus, low mass ratio evolution after rotational fission is responsible for creating synchronous binaries, high-$e$ binaries, asteroid pairs, re-shaped primaries, and potentially ternaries.

\clearpage

\subsection{Asteroid Pairs}
\label{sec:asteroidpairs}

YORP induced fission is a significant source of asteroid pairs~\citep{Pravec2010}. Simulated disrupted systems also escape with low escape velocities similar to those modeled for asteroid pairs. The primary spin periods of observed asteroid pairs have a very characteristic dependance on mass ratio. This dependence is captured naturally by the rotational fission process as shown in Fig. 15--the dots are the primaries of simulated disrupted systems and the crosses are observed asteroids.

[Figure 15]

\section{Discussion}
\label{discussion}

This is not the first rotational fission model for asteroid binary systems~\citep{Pravec2007, Scheeres2007b, Walsh2008b, Holsapple2010}, but this rotational fission model explains all observed NEA systems and constructs the entire life history of NEA systems into one coherent theory. This theory agrees with previous authors that the progenitors of the NEA population are disrupted critically spinning asteroids, and that the YORP effect forces asteroids through an effective random walk up to that disruption limit, or away towards the slow rotator population, although eventually slow rotators may be spun back up in the other direction or tumbling might prevent this.

Unlike previously presented rotational fission theories for binary asteroids, this work modeled the evolution of disrupted systems over long timescales and concluded that these systems always disrupt. In order to prevent disruption, these systems need to transfer energy out of the orbit of the system into the spin energy of the bodies, either to stay or be dissipated. In the Walsh et al. (2008) model the energy is dissipated via accretion of more material onto the secondary. This material is from subsequent rotational fission of the primary after the initial rotational fission. However, we discover that rotational fissioned systems disrupt on timescales longer ($\sim10^2$ to $10^3$ orbits) than the Walsh et al. (2008) model allows the system to evolve ($\sim5$ orbits) before implementing an impulsive YORP torque on the primary, but on timescales much shorter than the equivalent natural YORP torque would take to develop ($\gtrsim10^6$ orbits, estimated from the YORP timescale). 

The dynamical model is scale independent and so predicts that binary formation occurs at the same rate across the entire size distribution of NEAs, but other non-incorporated effects begin to play an important role. Bodies $< 100$ m in radius may be dominated by cohesive forces so the ``rubble pile'' approximation no longer applies~\citep{Scheeres2010b}. This would reproduce the observed disappearance of the spin barrier at small size scales~\citep{Pravec2007}. The YORP effect depends on the radius of the body to the second power, so at large sizes ($>10$ km) the timescale of the YORP effect approaches the NEA lifetime. These effects create a range of sizes for which we expect binary asteroids to be formed from rotational fission in: $0.1$ to $10$ km. This agrees well with the sizes of the observed binary NEA population--between $0.3$ and $3$ km~\cite{Pravec2007}, although the upper limit may also be set by small number statistics rather than the YORP timescale.

This formation mechanism predicts asteroid pairs amongst the NEA population. These will be harder to detect than their counterparts amongst the small Main Belt asteroid population, since the orbit scattering time is much shorter due to interactions with the inner planets, and since the progenitors are typically smaller, the secondary member of each pair will have a small absolute size ($<100$ m), which makes orbit determination difficult. In the small Main Belt asteroid pair population, this theory agrees with the already accepted idea that each pair is formed in a rotational fission event~\cite{Pravec2010}.

Shortcomings of the above model fall into two camps: unknown parameters and computational shortcuts. Geophysical parameters including the tidal Love number and the tidal quality number have significant uncertainties. In the model above, we were forced to assume a flat initial mass ratio, essentially the internal component mass distribution, since the actual distribution is unknown. Computationally, we took a number of shortcuts to reduce complexity and computational time including very simple impact physics, first-order tidal models, and second-order gravity. These assumptions probably had only a small impact on the efficiencies in the code. The largest computational time saver was the assumption of a planar system--an assumption supported by the high angular momentum content of rotationally fissioned asteroids, but this assumption removes an energy dissipation mechanism. When these systems fission, the components will most likely not be rotating about their principal axes. Thus, each component may damp energy through internal torques induced from non-prinicipal axis rotation. This model has the ability to be improved with more observations and more complete physics, however we feel that none will change the overall conclusions regarding the evolutionary tracks, but they will have an impact on the efficiencies of the different pathways.

\section{Conclusion}
\label{conclusion}

The evolution of NEA systems is driven by four important processes: initial rotational fission, secondary fission, impacts, and solar gravitational perturbations. The lower the mass ratio, the faster the spin rate required for initial rotational fission, and thus the more energy in the eventual binary system. The free energy transitions from positive to negative at a mass ratio of $0.2$ for the spherical end state, this divides the evolution of rotationally fissioned systems into two paths as shown in Fig. 1. Secondary fission can occur before low mass ratio systems are ejected. Enough energy is transferred to the secondary via spin-orbit coupling so that it undergoes rotational fission and creates a chaotic ternary as shown in Fig. 2. Secondary fission grows increasingly likely as mass ratio decreases, since the initial energy in the system increases and rotational energy transferred to the secondary is more effective on a less massive secondary. Chaotic ternaries are formed from secondary fission and evolve quickly back into a chaotic binary state, however impacts dissipate energy and produce more stable binaries. Escape of ternary members can also stabilize the system. Solar gravitational perturbations are important in changing the eccentricity and are responsible for both stabilizing and destabilizing binary systems. NEAs are actively evolving systems driven by these four processes and the observed asteroid classes are stages in this evolution.

Radiative processes dominate the evolution of the NEA population from the Yarkovsky effect which drives small Main Belt asteroids into resonances with Jupiter pushing them into the NEA population, to the YORP effect which dominates their spin evolution and forces them to disrupt forming asteroid systems, to the BYORP effect which drives these systems back together or apart. The lives of NEAs are exciting--each asteroid may go through many iterations of the cycle shown in Fig. 1 taking different paths each time.

\appendix
\label{sec:appendix}

\section{Derivation of the Two-Body Equations of Motion}
\label{sec:derivationtwobody}
The equations of motion of the two-body system will be derived from the Euler-Lagrange equations of motion modified to account for mutual body tides and then placed in the rotating coordinate frame of a body encircling the Sun at 1 AU to account for solar tides (using the Hill approximation).

The two-body integrator models the system as two tri-axial ellipsoids, $E_1$ and $E_2$, expanded to 2nd order in their moments of inertia. A relative coordinate system with four degrees of freedom is defined: $r$ is the separation distance between the centers of mass of the two bodies, $\theta$ tracks the rotation of the line connecting the centers of mass relative to an inertial frame, and $\phi_n$ tracks the rotation of the body $n$ with respect to the line connecting the centers of mass. Figure A.16 shows a schematic of the two-body coordinate system. In order to track the rotation of a body in an inertial frame, $\psi_n$, the two coordinates would need to be added $\psi_n = \theta + \phi_n$.

[Figure A16]

The density of each body is $\rho = $ 2 g/cc--a typical density for small bodies in the solar system. Each tri-axial ellipsoid is a prolate body with axes $\alpha_n > \beta_n = \gamma_n$, where $\gamma_n$ is oriented along the rotation axis. All rotation axes are aligned, thus all motion is constrained to a plane. 

The kinetic energy of the system $T$ has four independent degrees of freedom when written in the relative coordinate system:

\begin{equation}
T = \frac{1}{2} m \left( \dot{r}^2 + r^2 \dot{\theta}^2 \right) + \frac{1}{2} \sum_{n=1}^2 I_{n_z} \left( \dot{\phi}_n + \dot{\theta} \right)^2
\end{equation}

\noindent where $m$ is the reduced mass of the system and $I_{n_i}$ is the moment of inertia of body $n$ along axis $i$.

The potential energy $V$ used is a 2nd order expansion in the moments of inertia corresponding to tri-axial ellipsoids, and has three independent degrees of freedom when written in the relative coordinate system: 

\begin{multline} 
V = - \frac{G M_1 M_2}{r} \left\{ 1 + \frac{1}{2 r^2} \left[ \bar{I}_1 + \bar{I}_2  - \frac{3}{2} \left( \bar{I}_{1_x} + \bar{I}_{1_y} + \bar{I}_{2_x} + \bar{I}_{2_y} \right. \right. \right. \\
 \left. \left. \left. - \cos \left(2 \phi _1\right) \left( \bar{I}_{1_y} - \bar{I}_{1_x} \right)  - \cos \left(2 \phi _2 \right) \left( \bar{I}_{2_y} - \bar{I}_{2_x} \right) \right) \right] \right\}
\end{multline}

\noindent where $M_n$ is the mass of body $n$, $\bar{I}_n$ is the sum of the reduced (mass normalized) inertial moments of body $n$, and $\bar{I}_{n_i}$ is the reduced (mass normalized) moment of inertia of body $n$ along axis $i$. 

The modified Euler-Lagrange equations of motion for this system:

\begin{equation}
\frac{d}{d t} \left( \frac{\partial L}{\partial \dot{q}} \right) = \frac{\partial L}{\partial q} - \Gamma_q
\end{equation}

\noindent where the tidal torque term $\Gamma_q$ appears only in the relative spin coordinate equations as $\Gamma_{\phi_1}$ or $\Gamma_{\phi_2}$, and is zero for all others, $\Gamma_{r} = \Gamma_{\theta} = 0$. This tidal torque is responsible for dissipating energy in the form of heat. The exact form of the tidal torque is discussed in~\ref{sec:derivationtidaltheory}.

The Euler-Lagrange equation of motion for the separation distance $r$ is solved straightforwardly for $\ddot{r}$:

\begin{equation}
\ddot{r} = r \dot{\theta}^2 - \frac{V_r}{m}
\end{equation}

\noindent where $V_r$ is the partial derivative of the potential energy with respect to the separation distance $r$:

\begin{multline} 
V_r = \frac{G M_1 M_2}{r^2} \left( 1 + \frac{3}{2 r^2} \left( \bar{I}_1 + \bar{I}_2  - \frac{3}{2} \left( \bar{I}_{1_x} + \bar{I}_{1_y} + \bar{I}_{2_x} + \bar{I}_{2_y} \right. \right. \right. \\
 \left. \left. \left. - \cos \left(2 \phi _1\right) \left( \bar{I}_{1_y} - \bar{I}_{1_x} \right)  - \cos \left(2 \phi _2 \right) \left( \bar{I}_{2_y} - \bar{I}_{2_x} \right) \right) \right) \right)
\end{multline}

Since the Lagrangian does not depend directly on $\theta$, the right-hand side of the orbital Euler-Lagrange equation of motion is zero and so the it becomes a statement of the conservation of angular momentum:

\begin{equation}
\frac{d}{d t} \left( I_{1_z} \dot{\phi}_1 +  I_{2_z} \dot{\phi}_2 + I_z \dot{\theta} \right) = 0
\end{equation}

\noindent where $I_z = I_{1_z} + I_{2_z} + m r^2$ and is an abbreviation for the system or polar moment of inertia.

The modified Euler-Lagrange equations of motion for the relative spin coordinates $\phi_n$ for each body $n$:

\begin{equation}
I_{n_z} \ddot{\phi}_n + I_{n_z} \ddot{\theta} = - V_{\phi _n}  - \Gamma_n
\end{equation}

\noindent where $V_{\phi_n}$ is the partial derivative of the potential energy with respect to the relative spin coordinate $\phi_n$:

\begin{equation}
V_{\phi _n} =  \frac{3}{2} \left( \frac{G M_1 M_2}{r^3} \right)  \left( \bar{I}_{n_y} - \bar{I}_{n_x} \right) \sin \left( 2 \phi _n \right)
\end{equation}

The modified Euler-Lagrange equations of motion for the angular coordinates can be arranged in a matrix representation:

\begin{equation}
\left(
\begin{array}{ccc}
 I_{1_z} & 0 & I_{1_z} \\
 0 & I_{2_z} & I_{2_z} \\
 I_{1_z} & I_{2_z} & I_z
\end{array}
\right) 
\left(
\begin{array}{c}
 \ddot{\phi }_1 \\
 \ddot{\phi }_2 \\
 \ddot{\theta} 
\end{array}
\right)
=
\left(
\begin{array}{c}
- \Gamma _1 - V_{\phi _1} \\
- \Gamma _2 - V_{\phi _2} \\
 -2 m r \dot{r} \dot{\theta }
\end{array}
\right)
\end{equation}

Solving all three angular equations of motion simultaneously gives the equations of motion for the individual angular coordinates:

\begin{equation}
\ddot{\theta } = - \frac{2 \dot{r} \dot{\theta }}{r} + \frac{ \Gamma _1 + \Gamma _2 + V_{\phi _1} + V_{\phi _2} }{m r^2} 
\end{equation}
\begin{equation}
\ddot{\phi }_1 = \frac{2 \dot{r} \dot{\theta }}{r} - \frac{\Gamma _1 + \Gamma _2 + V_{\phi _1} + V_{\phi _2} }{m r^2} - \frac{ \Gamma _1 + V_{\phi _1}}{I_{1_z}} 
\end{equation}
\begin{equation}
\ddot{\phi }_2 = \frac{2 \dot{r} \dot{\theta }}{r} - \frac{ \Gamma _1 + \Gamma _2 + V_{\phi _1} + V_{\phi _2} }{m r^2} - \frac{ \Gamma _2 + V_{\phi _2}}{I_{2_z}} 
\end{equation}

These relative equations of motion are in the orbiting reference frame of the asteroid system. This system can be transformed to the inertial frame of the Sun via Hill's approximation; planar motion has already been assumed, but we also now assume a circular heliocentric orbit:

\begin{equation}
\ddot{r}_s  = \ddot{r} + 3 n^2 r \cos \theta + 2 n r \dot{\theta}
\end{equation}
\begin{equation}
\ddot{\theta}_s = \ddot{\theta} - 2 n \dot{r}
\end{equation}

\noindent where $n$ is the mean motion of the asteroid system about the Sun.Thus, the equations of motion for the two-body integrator are:

\begin{equation}
\ddot{r}_s = r \dot{\theta}^2 - \frac{V_r}{m} + 2 n r \dot{\theta} + 3 n^2 r \cos \theta
\end{equation}
\begin{equation}
\ddot{\theta}_s  =- \frac{2 \dot{r} \dot{\theta }}{r} + \frac{ \Gamma _1 + \Gamma _2 + V_{\phi _1} + V_{\phi _2} }{m r^2} - 2 n \dot{r}
\end{equation}
\begin{equation}
\ddot{\phi }_1  = \frac{2 \dot{r} \dot{\theta }}{r} - \frac{\Gamma _1 + \Gamma _2 + V_{\phi _1} + V_{\phi _2} }{m r^2} - \frac{ \Gamma _1 + V_{\phi _1}}{I_{1_z}} 
\end{equation}
\begin{equation}
\ddot{\phi }_2  = \frac{2 \dot{r} \dot{\theta }}{r} - \frac{ \Gamma _1 + \Gamma _2 + V_{\phi _1} + V_{\phi _2} }{m r^2} - \frac{ \Gamma _2 + V_{\phi _2}}{I_{2_z}} 
\end{equation}

\section{Derivation of the Three-Body Equations of Motion}
\label{sec:derivationthreebody}

The equations of motion of the three-body system will be derived from the Euler-Lagrange equations of motion modified to account for mutual body tides and then placed in the rotating coordinate frame of a body encircling the Sun at 1 AU to account for solar tides (Hill approximation). The mutual body tide between the two smallest members is neglected.

The three body integrator models the system as one tri-axial ellipsoid, $E_1$, and two spheres, $S_2$ and $S_3$. The system is described in an inertial cartesian coordinate system with nine coordinates. Each body $n$ has three coordinates: $x_n$ and $y_n$ track the body's center of mass and $\psi_n$ tracks the rotation angle. Figure B.17 shows a schematic of the three-body coordinate system.

[Figure B17]

The rotation $\psi_m$, orbital angle $\theta_{1m}$ and relative spin angles $\phi_{1m}$ of the sphere $m$ are shown schematically in Fig. B.18 and are related:

\begin{equation}
\phi_{1m} = \theta_{1m} - \psi_1 = \arctan \left( \frac{y_m - y_1}{x_m - x_1} \right) - \psi_1
\end{equation}

[Figure B18]

The density of each body is $\rho = $ 2 g/cc--a typical density for small bodies in the solar system. The tri-axial ellipsoid is a prolate body with axes $\alpha_1 > \beta_1 = \gamma_1$, where $\gamma_1$ is oriented along the rotation axis. The spheres are defined by a radius $R_m$. All rotation axes are aligned, thus all motion is constrained to a plane.

The kinetic energy of the system $T$ has nine independent degrees of freedom when written in the cartesian/angular coordinate system:

\begin{equation}
T = \frac{1}{2} \sum_{n=1}^3 M_n \left( \dot{x}_{n}^2 + \dot{y}_{n}^2 + \bar{I}_n \dot{\psi}_n^2 \right)
\end{equation}

\noindent where $\bar{I}_n$ is the sum of the reduced (mass normalized) inertial moments of body $n$.

The potential energy $V$ used for the primary is a 2nd order expansion in the moments of inertia corresponding to a tri-axial ellipsoid, and each of the secondaries is a Keplerian potential corresponding to a sphere. The potential energy has 7 independent degrees of freedom when written in the cartesian/angular coordinate system: 

\begin{multline} 
V =  - \frac{G M_2 M_3}{r_{23}}\\
- \frac{G M_1 M_2}{r_{12}} \left\{ 1 + \frac{1}{2 r_{12}^2} \left[ \bar{I}_1 - \frac{3}{2} \left( \bar{I}_{1_x} + \bar{I}_{1_y} - \cos \left(2 \phi _{12} \right) \left( \bar{I}_{1_y} -\bar{I}_{1_x} \right) \right) \right] \right\} \\
- \frac{G M_1 M_3}{r_{13}} \left\{ 1 + \frac{1}{2 r_{13}^2} \left[ \bar{I}_1 - \frac{3}{2} \left( \bar{I}_{1_x} + \bar{I}_{1_y} - \cos \left(2 \phi _{13} \right) \left( \bar{I}_{1_y} - \bar{I}_{1_x} \right) \right) \right] \right\}
\end{multline}

\noindent where $\bar{I}_{n_i}$ is the reduced (mass normalized) moment of inertia of body $n$ along axis $i$ and $r_{nm}$ is:

\begin{equation}
r_{nm} = \sqrt{ \left( x_n - x_m \right)^2 + \left( y_n - y_m \right)^2 }
\end{equation}

The Euler-Lagrange equations of motion for the cartesian coordinates are simply solved for $\ddot{x}$ and $\ddot{y}$:

\begin{equation}
\ddot{x}_n = \frac{1}{M_n} \frac{\partial V}{\partial x_n} \qquad \ddot{y}_n = \frac{1}{M_n} \frac{\partial V}{\partial y_n}
\end{equation}

The modified Euler-Lagrange equation of motion for the angular coordinate of the primary can be solved for the $\ddot{\psi}_1$:

\begin{equation}
\ddot{\psi}_1 = \frac{1}{I_1} \left( \frac{d V}{d \psi_1} - \Gamma_{21} - \Gamma_{31} \right)
\end{equation}

\noindent There are two tidal torques, $\Gamma_{m1}$, from each spherical body $m$ onto the primary. These tidal torques are responsible for dissipating energy in the form of heat. The tidal torque depends linearly on the relative spin angle rate, $\dot{\phi}_{1m}$, and inversely on the distance between the bodies, $r_{1m}$ to the fifth power. The exact form of the tidal torque is discussed in Section~\ref{sec:derivationtidaltheory}. 

The modified Euler-Lagrange equations of motion for the angular coordinates of the spheres are solved:

\begin{equation}
\ddot{\psi}_m = \frac{1}{I_m} \Gamma_{1m}
\end{equation}

\noindent where there is only one tidal torque, $\Gamma_{1m}$, acting on each spherical body from the primary. The tidal torque between the spherical bodies is neglected. These tidal torques are responsible for dissipating energy in the form of heat. The tidal torque depends linearly on the relative spin angle rate, $\dot{\phi}_{m}$, and inversely on the distance between the bodies, $r_{1m}$ to the fifth power. The exact form of the tidal torque is discussed in Section~\ref{sec:derivationtidaltheory}.

These relative equations of motion are in the orbiting reference frame of the asteroid system. This system can be transformed to the inertial frame of the Sun via Hill's approximation; planar motion has already been assumed, but we also now assume a circular heliocentric orbit:

\begin{equation}
\ddot{x}_{ns} = \ddot{x}_n + 2 n \dot{y}_n + 3 n^2 \dot{x}_n
\end{equation}
\begin{equation}
\ddot{y}_{ns} = \ddot{y}_n - 2 n \dot{x}_n
\end{equation}
\begin{equation}
\ddot{\psi}_{ns} = \ddot{\psi}_{n} 
\end{equation}

\noindent where $n$ is the mean motion of the asteroid system about the Sun.Thus, the equations of motion for the two-body integrator are:

\begin{equation}
\ddot{x}_{ns} = \frac{1}{M_n} \frac{\partial V}{\partial x_n} + 2 n \dot{y}_n + 3 n^2 \dot{x}_n
\end{equation}
\begin{equation}
\ddot{y}_{ns} = \frac{1}{M_n} \frac{\partial V}{\partial y_n} - 2 n \dot{x}_n
\end{equation}
\begin{equation}
\ddot{\psi}_{1s} = \frac{1}{I_1} \left( \frac{d V}{d \psi_1} - \Gamma_{21} - \Gamma_{31} \right)
\end{equation}
\begin{equation}
\ddot{\psi}_{ms} = \frac{1}{I_m} \Gamma_{1m}
\end{equation}

\section{Derivation of the Tidal Theory}
\label{sec:derivationtidaltheory}

The model applies the classical tidal torque presented in~\citet{Murray1999} for a spherical (point source) satellite $j$ acting on a spherical body $i$:

\begin{equation}
\Gamma_{i} = \text{sign}\left( \dot{\phi}_i \right) \frac{3}{2} k \left( \frac{3}{4 \pi  \rho_i } \right)^2 \frac{G M_i^2 M_j^2}{r_{ij}^6 R_i} \sin (2 \epsilon_i ) 
\end{equation}

\noindent  where $k$ is the tidal Love number and $\epsilon$ is the tidal lag angle. $R_i$ is the radius of body $i$ if it were a sphere of equal mass. The tidal bulge is independent of the shape of the body. The sign of $\dot{\phi}_i$ determines whether the tidal bulge is leading or trailing the tide-raising satellite, which determines the direction of angular momentum transfer between the orbit and the spin state. The tidal lag angle can be related to the specific tidal dissipation function $Q$, which describes how effective the body is at tidally dissipating energy:

\begin{equation}
Q = \frac{1}{\tan 2 \epsilon} \approx \frac{1}{2 \epsilon}
\end{equation}

However this classical torque presents a problem when $\dot{\phi}_i$ changes through zero, which occurs for many of these systems due to the chaotic nature of their evolution and the large spin-orbit coupling. When $\dot{\phi}_i$ crosses zero, $\Gamma_i$ changes sign instantaneously. This is unphysical since the bulge is a real phenomenon and would have some finite crossing time. Instantaneous switching is a difficulty for numerical integration as well. We introduce a modified torque that will linearize $\Gamma_i$ when $\dot{\phi}_i \approx 0$. 

\begin{equation}
\begin{centering}
\Gamma_i =
\left\{\begin{matrix}
\Gamma_i &  | \dot{\phi}_i | > \delta_i \\
\Gamma_i \frac{\dot{\phi}_i}{\delta_i} & | \dot{\phi}_i | \leq \delta_i
\end{matrix}\right.
\end{centering}
\end{equation}

\noindent where $\delta_i$ is some small characteristic angular spin rate for body $i$. We can derive an appropriate small characteristic angular spin rate from the torque equation $\ddot{\phi }_i = \Gamma_i/I_{i_z}$:

\begin{equation}
\delta_i = \Delta \dot{\phi}_i =\frac{\Gamma_i}{I_{i_z}} \Delta t
\end{equation}

\noindent where $\Delta t$ is some characteristic time, which can be derived from the crossing time of a pressure (seismic) wave: 

\begin{equation}
\Delta t = \frac{\Delta l }{ \Delta v}
\end{equation}

\noindent where $\Delta l  = 2 R_i$ is the characteristic length scale of the body, and $\Delta v$ is the pressure (seismic) wave velocity $c_i$. The pressure wave velocity can be found from the central pressure $P_i$:

\begin{equation}
c_i = \sqrt{\frac{P_i}{\rho_i}} = \sqrt{\frac{2 \pi G \rho_i}{3}} R_i
\end{equation}

The small characteristic angular spin rate $\delta_i$ is now determined:

\begin{equation}
\begin{centering}
\delta_i = \frac{\Gamma_i}{I_{i_z}} \frac{2 R_i}{c_i} = \frac{\Gamma_i}{I_{i_z}} \sqrt{\frac{6}{\pi G \rho_i}}
\end{centering}
\end{equation}

This gives the modified tidal torque:

\begin{equation}
\Gamma_i =
\left\{\begin{matrix}
\text{sign}\left( \dot{\phi}_i \right) \frac{3}{2} k \left( \frac{3}{4 \pi  \rho_i } \right)^2 \frac{G M_i^2 M_j^2}{r_{ij}^6 R_i} \sin (2 \epsilon_i ) &  | \dot{\phi}_i | > \delta_i \\
\sqrt{\frac{\pi G \rho_i}{6}} I_{i_z} \dot{\phi_i} & | \dot{\phi}_i | \leq \delta_i
\end{matrix}\right.
\end{equation}

\section{Secondary Fission Condition}
\label{sec:derivationsecondaryfission}

The coordinate system is given in Fig. D.19, and the condition for secondary fission is:

[Figure D19]

\begin{equation}
\ddot{\vec{r}}_{2B}^R \cdot \hat{r}_{2B} > 0
\end{equation}

\begin{multline}
\ddot{\vec{r}}_{2B}^R \cdot \hat{r}_{2B} =  - \frac{G M_A}{|\vec{r}_{AB}|^2} + |\vec{r}_{2B}| |\Omega_2|^2 - \left( 1 + \frac{M_B}{M_A} \right)^{-1} \frac{G M_1}{|\vec{r}_{1B}|^2} \left[ 1 + \frac{3}{2 |\vec{r}_{1B}|^2} \times \right.  \\
\left. \left( \bar{I}_1 - \frac{3}{2} \left( \bar{I}_{1x} + \bar{I}_{1y} + ( \bar{I}_{1x} -  \bar{I}_{1y} ) \cos 2 \psi_{1B} \right) \right) \right]  \left( \hat{r}_{1B} \cdot \hat{r}_{2B} \right) - \frac{G M_1}{|\vec{r}_{1A}|^2} \times \\
\left[ 1 + \frac{3}{2 |\vec{r}_{1A}|^2} \left( \bar{I}_1 - \frac{3}{2} \left( \bar{I}_{1x} + \bar{I}_{1y} + ( \bar{I}_{1x} -  \bar{I}_{1y} ) \cos 2 \psi_{1A} \right) \right) \right]  \left( \hat{r}_{1A} \cdot \hat{r}_{2B} \right) \\
\end{multline}

\section{Impact Mass Redistribution}
\label{sec:impactmassredistribution}

When the spherical secondary re-impacts the tri-axial primary, the mass of the secondary is placed onto the primary so as to bring the surface closer to a geopotential  along the equator (oblate spheriod). The smaller $\beta$ axis is increased by a height:

\begin{equation}
h = \frac{R_s^3}{\alpha \beta}
\end{equation}

where $R_s$ is the radius of the secondary and $\alpha$ and $\beta$ are the original tri-axial axes.

\section{Modeling Tidal Timescales}
\label{sec:modelingtidaltimescales}

High mass ratio systems (mass ratio $>0.2$) were evolved from YORP induced rotational fission according to the two-body integrator described exactly as above with the exception of solar gravitational perturbations which were neglected for this case. According to theory developed in~\citet{Scheeres2009b}, if the secondary does not undergo secondary fission and without the influence of external events, then these systems will evolve from their initial orbital state immediately after rotational fission to the doubly synchronous (relative equilibrium) state with the same angular momentum. Mutual tidal dissipation naturally weakens and eventually turns off as the system approaches and reaches the doubly synchronous state.

The doubly synchronous state has lower energy than the initial state, and this energy is dissipated via mutual body tides. After evolving each system for $10^4$ years, some of the systems had reached the doubly synchronous state, and the time that the energy dissipation rate went effectively to zero was recorded as the time the system became a doubly synchronous system. Many systems had not reached the doubly synchronous state within $10^4$ years, due to limited computation resources the timescale for transformation to the doubly synchronous state was determined by extrapolation. The energy dissipation as a function of time for each system could be fit with a power law (with greater than 99\% confidence in the fit) and that power law was than extrapolated to zero when the system would be in the doubly synchronous state and there would be no more mutual tidal energy dissipation.

\section{Limit on the Mass Ratio of Next Secondary Fission}
\label{sec:massratiosecondaryfission}

Let there be a body made of two components $M_1$ and $M_2$. The first component is made up of two components $M_{11}$ and $M_{12}$, so that $M_{11} + M_{12} = M_1$. Without loss of generality assume that:
\begin{equation}
M_{12} > M_{11}
\end{equation}
In order that the $M_2$ component fission first the following condition must be met:
\begin{equation}
M_2 > M_{12} > M_{11}
\end{equation}
Divide by $M_1$ to put in terms of mass ratio.
\begin{equation}
\frac{M_2}{M_1} = q_2 > \frac{M_{12}}{M_1}
\end{equation}
\begin{equation}
\frac{1}{q_2} < \frac{M_{11}}{M_{12}} + 1 = \frac{1}{q_{12}} + 1
\end{equation}
\begin{equation}
q_{12} < \frac{q_2}{1 - q_2}
\end{equation}
This condition is not a strong condition since it doesn't require the mass ratio of a future rotational fission event to decrease after an initial rotational fission event.

\section{Acknowledgements}
\label{Acknowledgements}

The authors would like to thank Matija {\'C}uk and the anonymous reviewer for their timely and much appreciated feedback. The authors would like to acknowledge financial support from the NASA Earth and Space Science Fellowship, from NASA Grants NNX08AL51G from the Planetary Geology and Geophysics Program and NNX09AU23G from NASA's Outer Planets Research Program. 

\bibliographystyle{model2-names.bst}
\bibliography{bibliography.bib}

\begin{table*}[h!]
\begin{center}
\begin{tabular}{| p{22mm} | p{50mm} | p{54mm} |}
\hline
Morphology & Observed Examples & Description \\
\hline
\hline
Doubly Synchronous Binaries & Frostia, Hermes, and Gavrilin~\citep{Behrend2006, Margot2003, Higgins2008} & All spin rotation periods are equivalent to the orbital revolution period. Mass ratios $>0.2$.\\
\hline
Contact  Binaries & Castalia, Bacchus, and 2005 CR$_{37}$~ \citep{Hudson1994, Benner1999, Benner2006} & Single asteroids with a bi-modal shape appear as two similar-sized components resting on each other. Component mass ratios $>0.2$.\\
\hline
Synchronous Binaries & 1999 KW$_4$, 2000 DP$_{107}$, and 2002 CE$_{26}$~ \citep{Ostro2006, Margot2002, Shepard2006} & Secondary spin rotation period is equivalent to the orbital revolution period. Primary is fast rotating and has a characteristic oblate shape with an equatorial bulge. Mass ratios $<0.2$.\\
\hline
High-$e$ Binaries &  2004 DC and 2003 YT$_1$~ \citep{Taylor2008, Nolan2004} & Mutual orbit is eccentric and secondaries may not be synchronous. Otherwise resemble synchronous binaries. Mass ratios $<0.2$.\\
\hline
Ternaries & 2001 SN$_{263}$ and 1994 CC~ \citep{Nolan2008, Brozovic2009} & All three members are in the same plane, and the primary resembles the primary of the synchronous binary. Mass ratios $<0.2$.\\
\hline
Re-shaped Asteroids & 1999 RQ$_{36}$ and 2008 EV$_5$~ \citep{Nolan2007, Busch2010} & Single asteroids that  resemble the primary of the synchronous binary. \\
\hline
\end{tabular}
\caption{Examples of each observed NEA class according to a morphological classification scheme.}
\label{tab:asteroidtypes}
\end{center}
\end{table*}

\begin{table}[t]
\begin{center}
\begin{tabular}{| m{20mm} | c | c | c | c | c |}
\hline
Mass \ \ \ \ \ \ Ratio Bins & 0 - 0.05 & 0.05 - 0.1 & 0.1 - 0.15 & 0.15 - 0.2 \\
\hline
Time [Days] & $26_{14}^{56}$ & $39_{20}^{82}$ & $59_{37}^{133}$ & $197_{98}^{448}$ \\
\hline
\hline
Primary Shape Ratio Bins & 0.6 - 0.7 & 0.7 - 0.8 & 0.8 - 0.9 & 0.9 - 1.0 \\
\hline
Time [Days] & $35_{18}^{65}$ & $63_{36}^{144}$ & $104_{56}^{197}$ & $532_{270}^{1758}$ \\
\hline
\hline
Secondary Shape Ratio Bins & 0.6 - 0.7 & 0.7 - 0.8 & 0.8 - 0.9 & 0.9 - 1.0 \\
\hline
Time [Days] & $59_{28}^{132}$ & $54_{24}^{163}$ & $45_{23}^{127}$ & $39_{19}^{91}$ \\
\hline
\end{tabular}
\caption{The median disruption time (in days) for four bins of width 0.05 in mass ratio and 0.1 in shape ratio are tabulated. In the subscript is the 25th percentile and in the superscript is the 75th percentile of the data from each bin.}
\label{tab:escapetrends}
\end{center}
\end{table}

\begin{table}[t]
\begin{center}
\begin{tabular}{| m{17mm} | c | c |}
\hline
Impact &  Perpendicular & Tangential \\
Type & Velocity [cm/s] & Velocity [cm/s] \\
\hline
\hline
Primary Impact & $13_{7}^{22}$  & $59_{51}^{66}$ \\
\hline
Secondary Impact  & $9_{4}^{18}$ & $7_{4}^{13}$ \\
\hline
Binary\ \ \ \  Impact &  $3_{2}^{6}$ &  $37_{35}^{41}$ \\
\hline
\end{tabular}
\caption{For each type of impact, the median impact velocities as well as the first and third quartiles are tabulated for the perpendicular and tangential directions relative to the center of mass of the impacting bodies in cm/s. Collisions between the primary and one of the secondaries of a ternary are shown in the first row, collisions between the secondaries of a ternary are shown in the second row, and collisions between the primary and secondary of a binary are shown in the third row.}
\label{tab:impacts}
\end{center}
\end{table}

\clearpage

Figure 1 Caption: Evolutionary tracks for an NEA. $q$ is the rotational fission component mass ratio. Arrows indicate the direction of evolution along with the process propelling the evolution and a typical timescale. Simple schematics show evolutionary states, an underline indicates an observed asteroid class. Stable ternaries are rare, and so their continued evolution is not described here, although it should be noted that ternaries may be formed via multiple primary fission events. It is important to note that the eventual outcomes are single asteroids (re-shaped asteroids, contact binaries, each member of asteroid pairs), so this evolutionary process represents a binary cycle. 

Figure 2 Caption: The upper right hand corner shows a motivating image of Itokawa taken by the Hayabusa spacecraft (Image courtesy of ISAS/JAXA). The cartoons document the low mass ratio evolutionary model. Solid lines indicate surfaces and dashed lines indicate ``rubble-pile'' internal substructure. The ``rubble pile'' asteroid evolves from the upper left to bottom left chaotic binary panel via a YORP-induced rotational fission event. A secondary fission event occurs between the lower left and lower middle panels creating a chaotic ternary system. One of the components impacts the primary as the system evolves from the lower middle to lower right panel forming a less energetic, more stable binary system.

Figure 3 Caption: The semi-major axis $a$ (solid line and left-hand, vertical axis) and the eccentricity $e$ (dotted line and right-hand, vertical axis) are shown as a function of time for both a 1999 KW$_4$-like system (top plot) and an Itokawa-like system (bottom plot).

Figure 4 Caption: The rotational periods of the primary (solid line) and the secondary (dotted line) are shown as a function of time for both a 1999 KW$_4$-like system (top plot) and an Itokawa-like system (bottom plot). The dashed line is the period ($\sim 2.33$ hours) for the surface disruption of a sphere of density $2$ g/cc.

Figure 5 Caption: The average separation distance between the binary members measured in primary radii after $100$ years of evolution as a function of mass ratio for $150$ systems. This simulation does not allow secondary fission or include solar gravity perturbations just evolves the system according to the interactions of two aspherical bodies. The dark line indicates the Hill radius ($80.5$ primary radii) of these systems at 1 AU, crossing this radius is equivalent to escape for the needs of this work.

Figure 6 Caption: The timescale for the tidal evolution to the doubly synchronous state as a function of mass ratio. The nominal ejection timescale from the NEA population is $10^7$ years. The black data points are the results of numerical modeling and the black curve is a power law fit to those points.

Figure 7 Caption: Each point shows the time after initial rotational fission for the simulated system to disrupt and each cross indicates a system that survived $1,000$ years without disrupting. These simulations include tri-axial gravitational potentials, mutual body tides and solar gravitational perturbations. In order from top to bottom, the time to system disruption is shown as a function of system mass ratio, primary shape ratio, and secondary shape ratio.

Figure 8 Caption: These are the same systems as in Fig. 7 that disrupted, however now those systems that underwent secondary, surface fission are distinguished. In the left hand plots, those system that go through surface fission before disruption are shown as crosses at the time of secondary fission, while those that did not secondary fission are shown as dots at the time of disruption. The right hand plots show the fraction of systems in each bin of width 0.025 for mass ratios and 0.05 for shape ratios that underwent secondary, surface fission. The fraction and errors were calculated using the Wilson Score Confidence Interval. From top to bottom, The time to fission or disruption and the fission fraction are shown as functions of mass ratio, primary shape ratio and secondary shape ratio.

Figure 9 Caption: The perpendicular and tangential velocities in cm/s for every occurence of all the three types of impacts occurring in the simulation: The dots represent primary impacts--the collisions that occur between the primary and one of the secondaries of a ternary system, the crosses represent secondary impacts--the collisions that occur between secondary and tertiary members of ternary systems, the triangles represent binary impacts--the collisions that occur between secondary and primary members of binary systems.

Figure 10 Caption: Chaotic ternary to binary loop via secondary fission and three ternary processes. Each process has a schematic as well as the percent likelihood a system will follow that path. All of these processes are dynamic and occur on timescales much shorter than a year.

Figure 11 Caption: The type of each impact is represented by the same symbols as in Fig. 9. The top plot shows the total velocity in cm/s of each impact comparing them to the critical disruption velocities for rubble piles shown as the dashed line~\citep{Stewart2009}. The bottom plot shows the specific energy in erg/g for each impact comparing them to derived catastrophic disruption thresholds: the dotted lines represent models that include material strength and self-gravity: the top line is~\citet{Benz1999} and the bottom is~\citet{Durda1998}, the dashed lines represent models that only include self-gravity: the top line is~\citet{Love1996} and the bottom is~\citet{Davis1995}, and the dot-dashed line is the self-potential energy of the target.

Figure 12 Caption: This shows the primary and secondary shape ratios of the stable binaries. The symbols are the same as in Fig. 13.

Figure 13 Caption: From top to bottom, the distribution of semi-major axes, primary rotation periods and secondary rotation periods are shown as a function of mass ratio. The triangles represent the $7$ binary systems out of $450$ that did not disrupt in the simulation that did not allow secondary fission. The crosses indicate the $41$ stable binaries out of $526$ that were outputs of the simulation that did allow secondary fission to occur. The dots indicate observed systems~\citep{Pravec2007}. The dashed line in the bottom two plots is the critical spin disruption limit (2.33 hour period) for a body with a density of 2 g/cc, however it only includes bodies $\gtrsim 100$ m in radius (i.e. secondaries with a mass ratio less than 0.001 for a 1 km primary). Bodies smaller than $200$ m in diameter may have internal strength or cohesive binding and so will require greater spin rates to disrupt.

Figure 14 Caption: The timescale for synchronization due to mutual body tides as a function of mass ratio are shown as dots for each modeled system. The dotted line indicates the nominal lifetime for an NEA asteroid system. The crosses are the analytical de-spinning timescales for known asteroid systems~\citep{Goldreich2009}.

Figure 15 Caption: The primary spin period of asteroid pairs as a function of mass ratio. The dots are the output from the simulation and the crosses are observed asteroid pairs~\citep{Pravec2010}.

Figure A16 Caption: Two-body Coordinate System

Figure B17 Caption: Three-body Coordinate System

Figure B18 Caption: Three-body Angular Coordinate System

Figure D19 Caption: New Relative Coordinate System

\clearpage
Figure 1:
\begin{center}
\includegraphics[width=150mm]{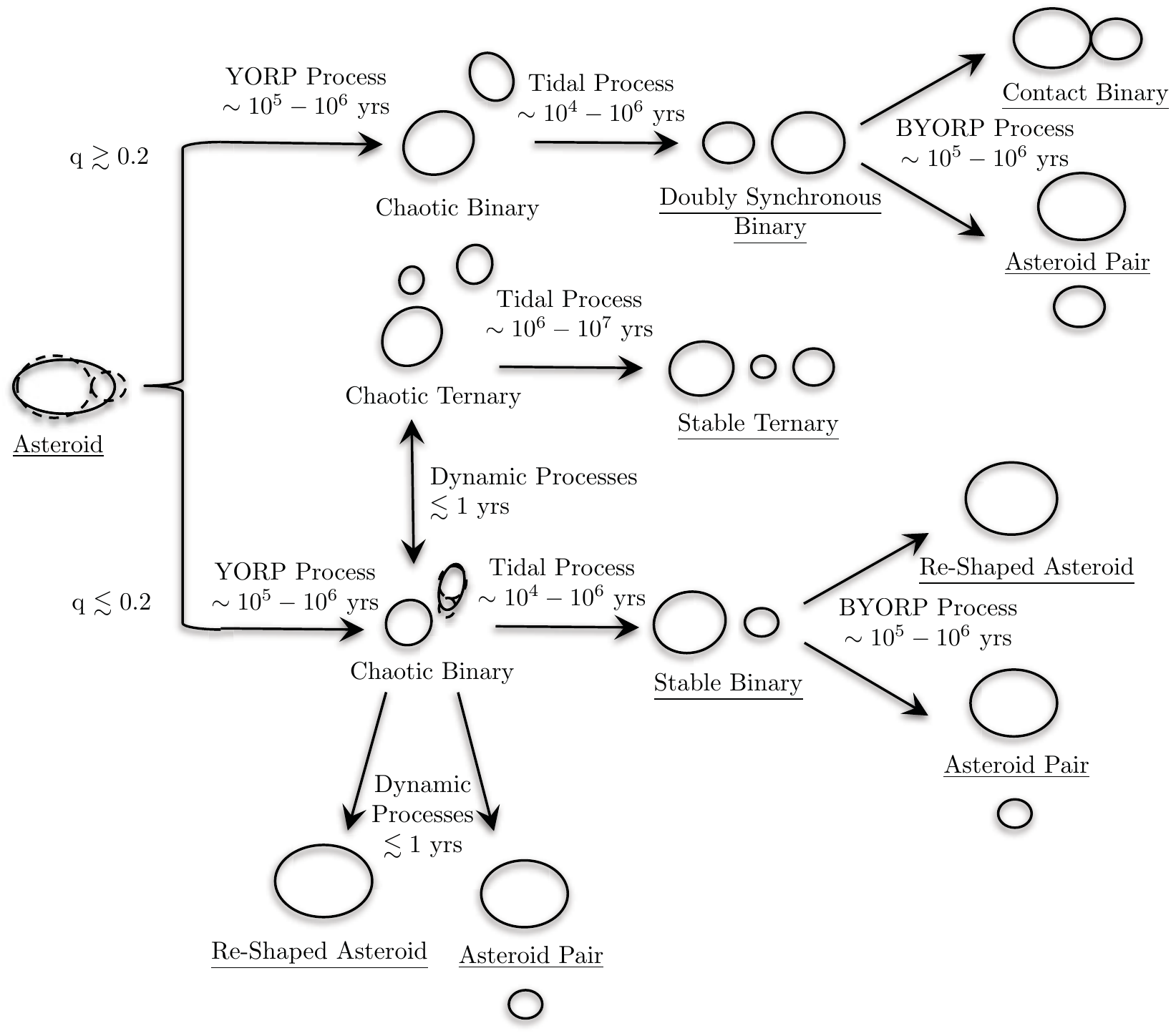}
\end{center}

\clearpage
Figure 2:
\begin{center}
\includegraphics[width= 90mm]{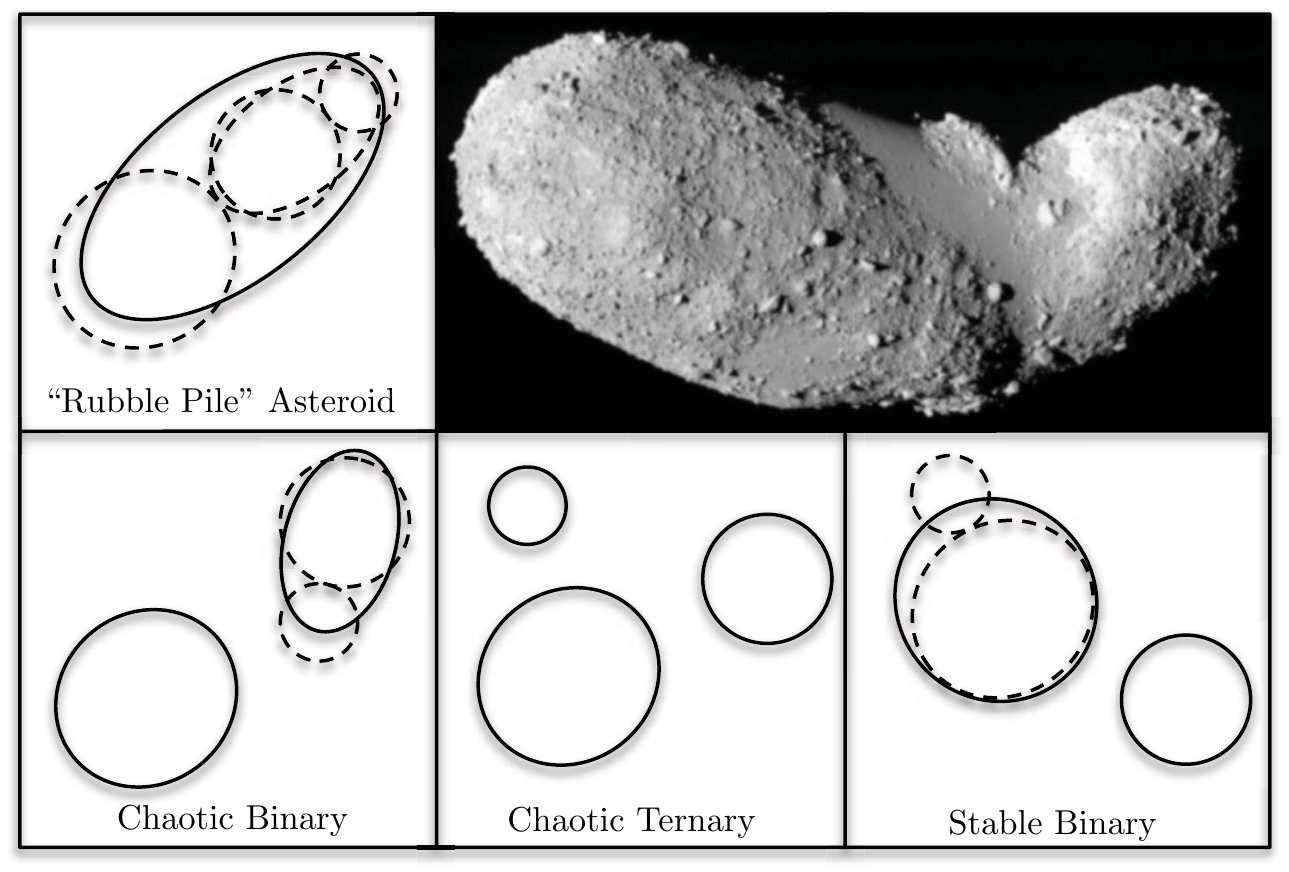}
\end{center}

\clearpage
Figure 3:
\begin{center}
\includegraphics[width= 90mm]{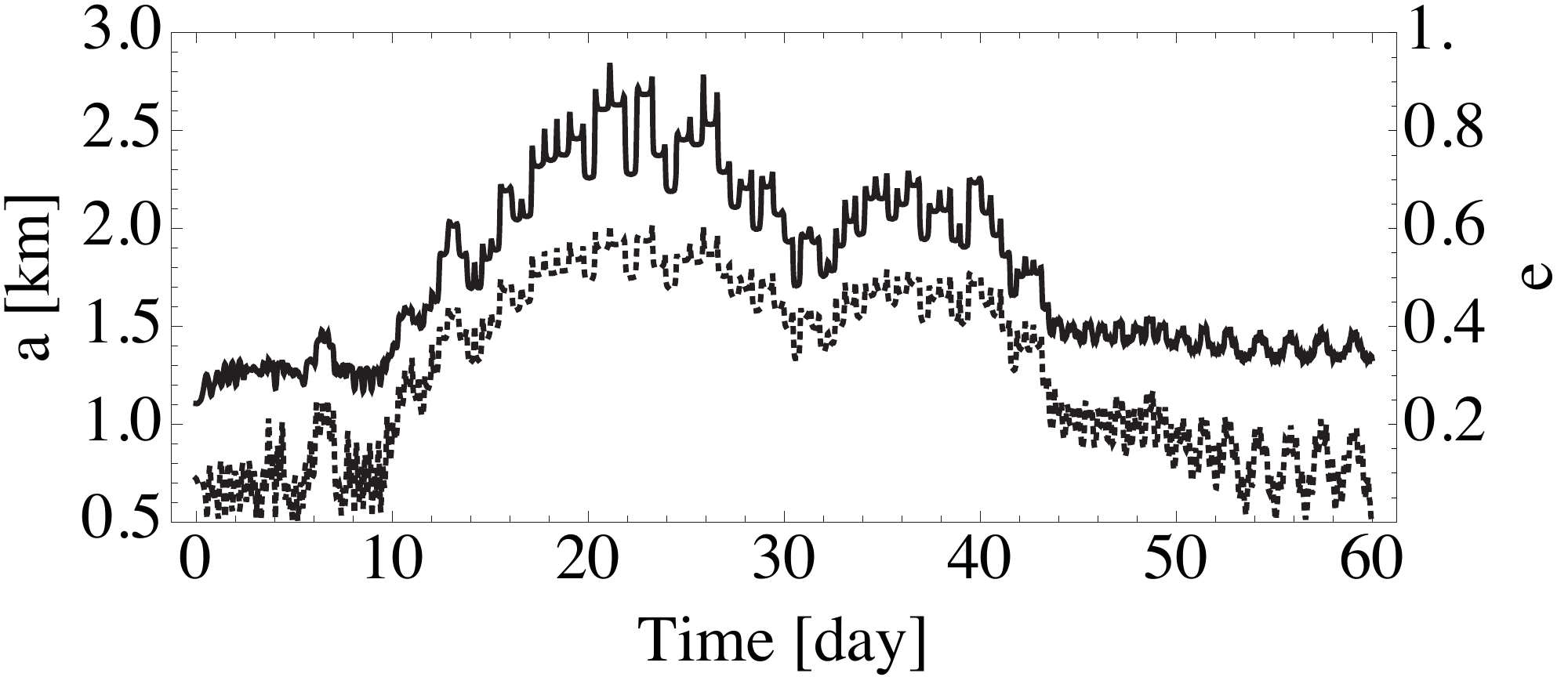}
\includegraphics[width= 90mm]{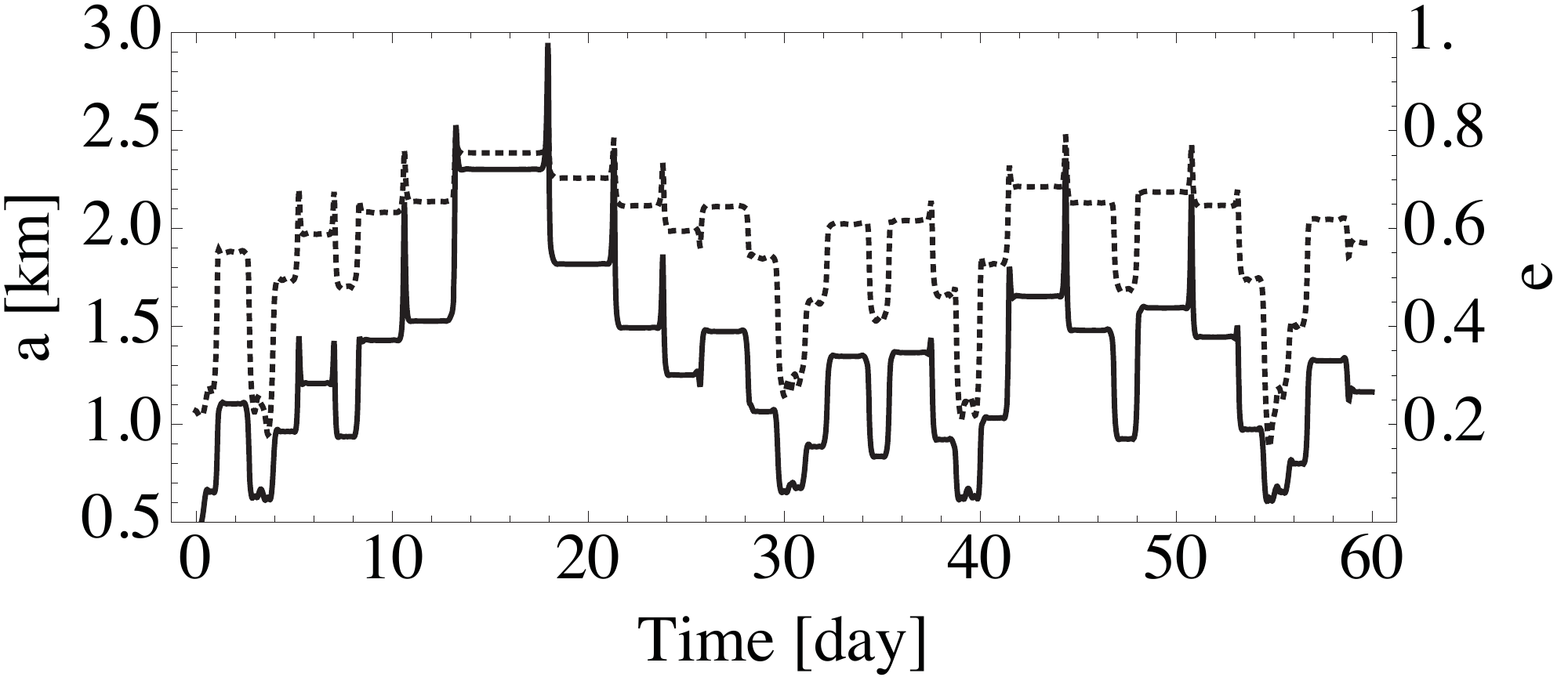}
\end{center}

\clearpage
Figure 4:
\begin{center}
\includegraphics[width= 90mm]{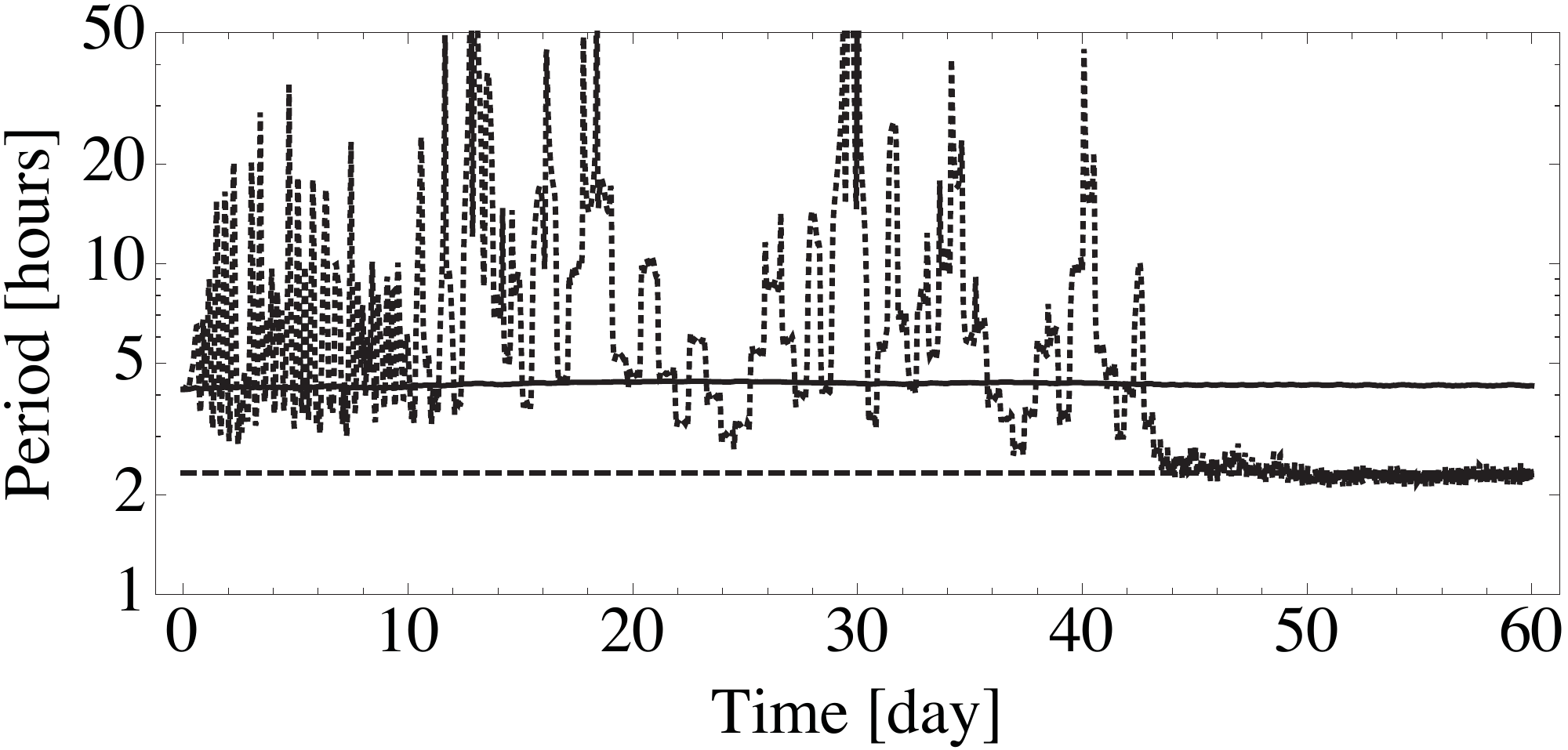}
\includegraphics[width= 90mm]{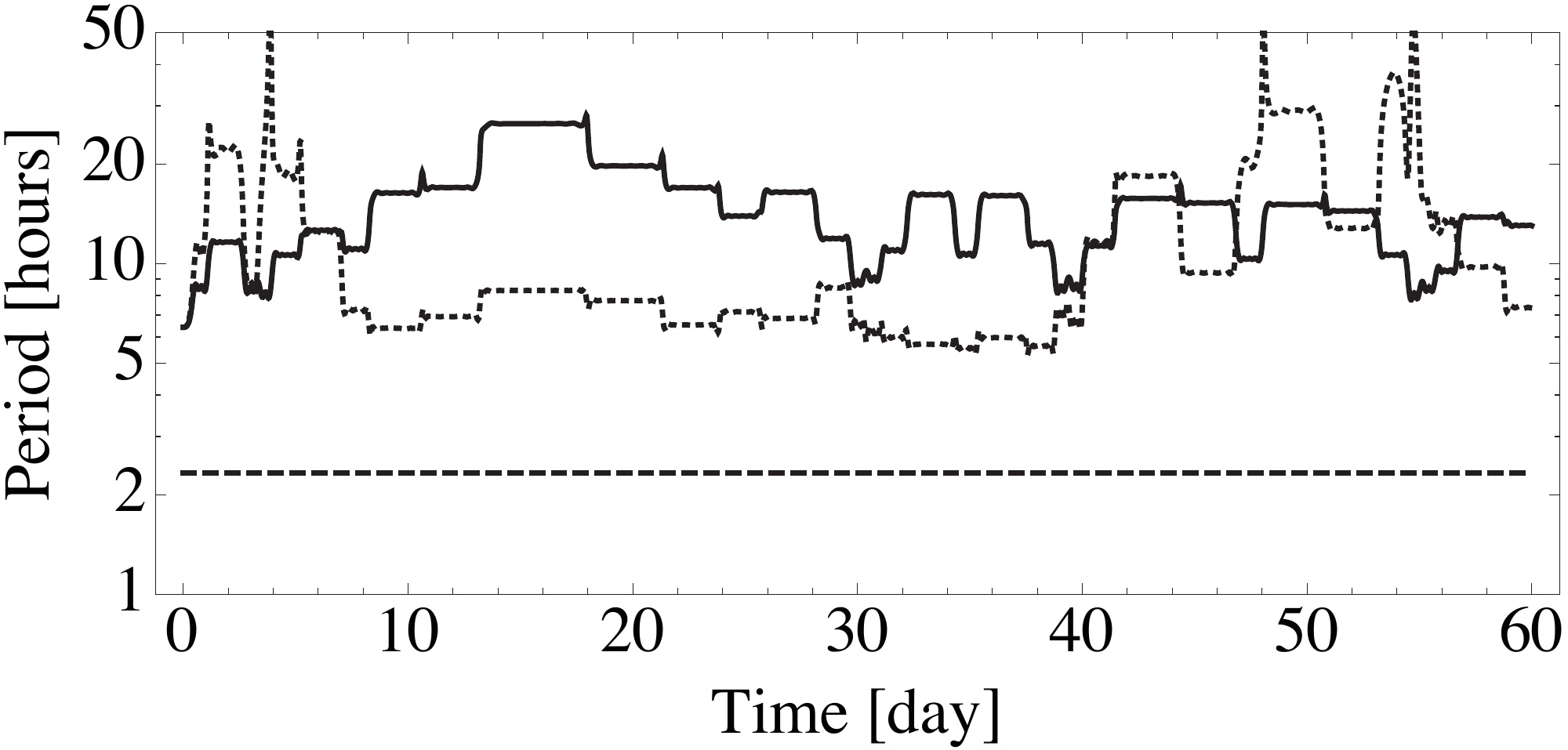}
\end{center}

\clearpage
Figure 5:
\begin{center}
\includegraphics[width= 90mm]{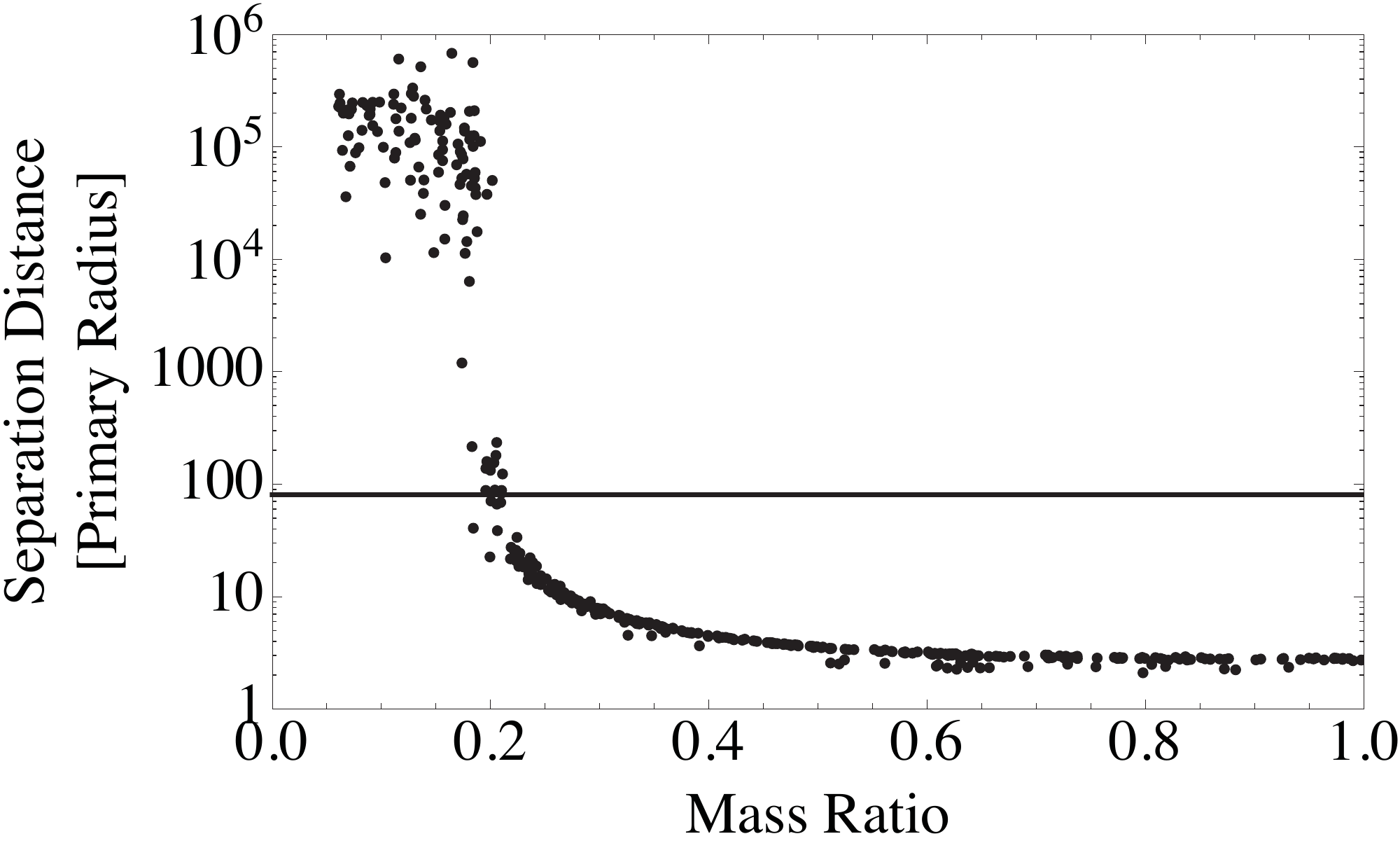}
\end{center}

\clearpage
Figure 6:
\begin{center}
\includegraphics[height= 90mm,angle=90]{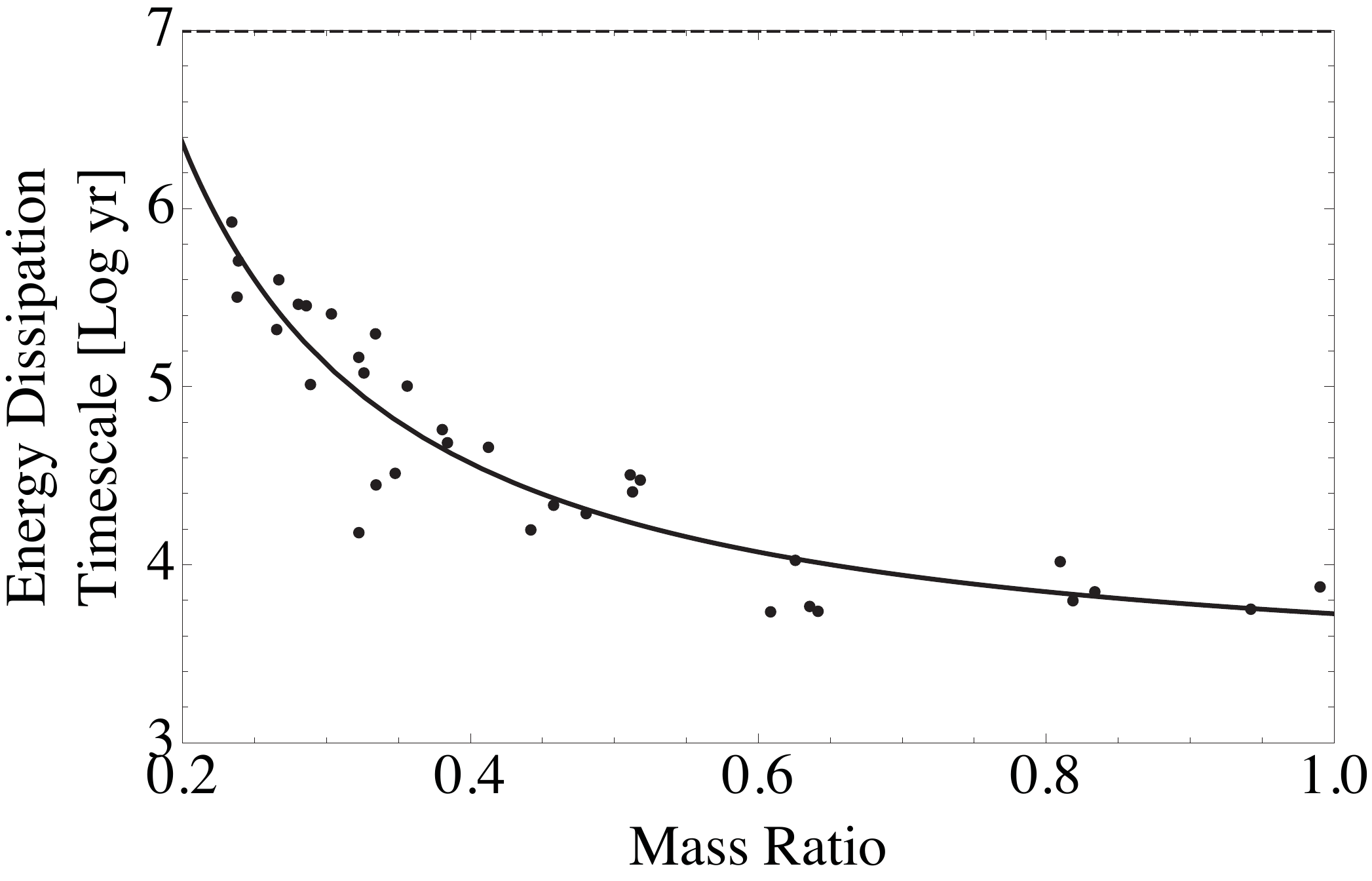}
\end{center}

\clearpage
Figure 7:
\begin{center}
\includegraphics[width=90mm]{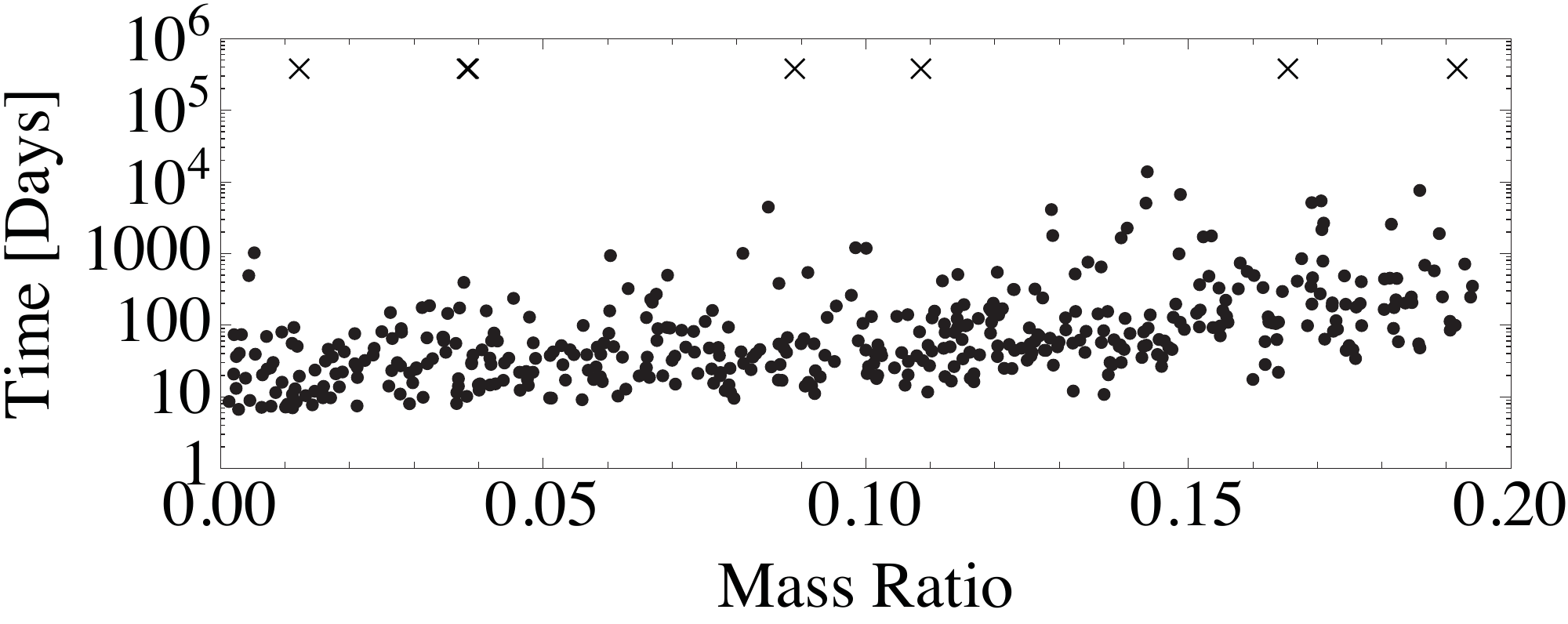}
\includegraphics[width=90mm]{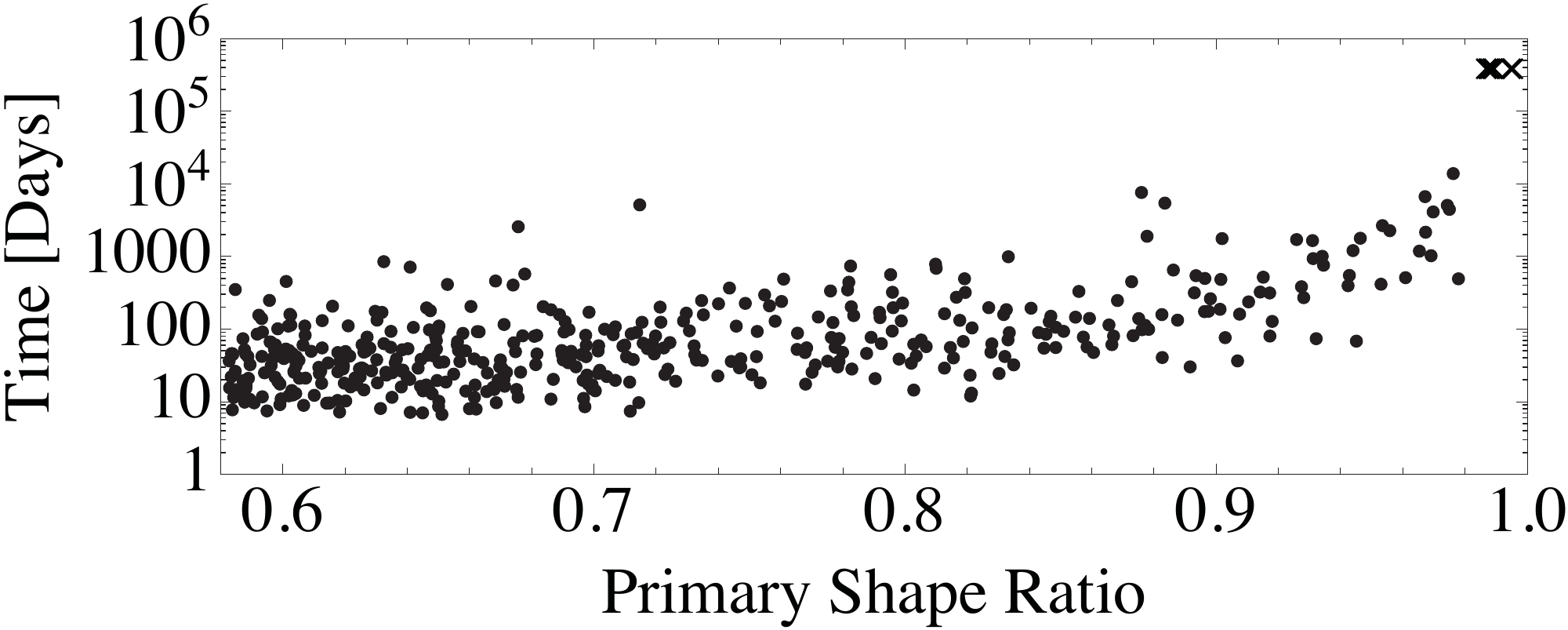}
\includegraphics[width=90mm]{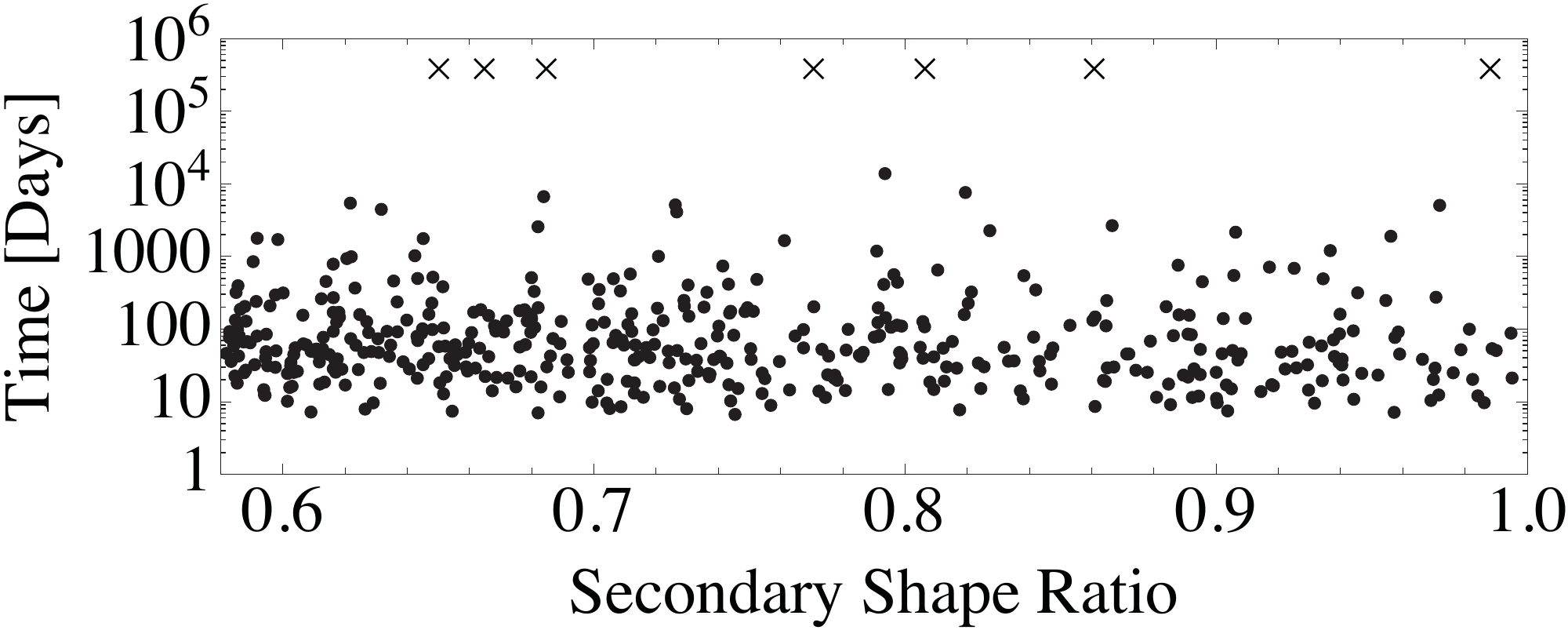}
\end{center}

\clearpage
Figure 8:
\begin{center}
\includegraphics[width=67mm]{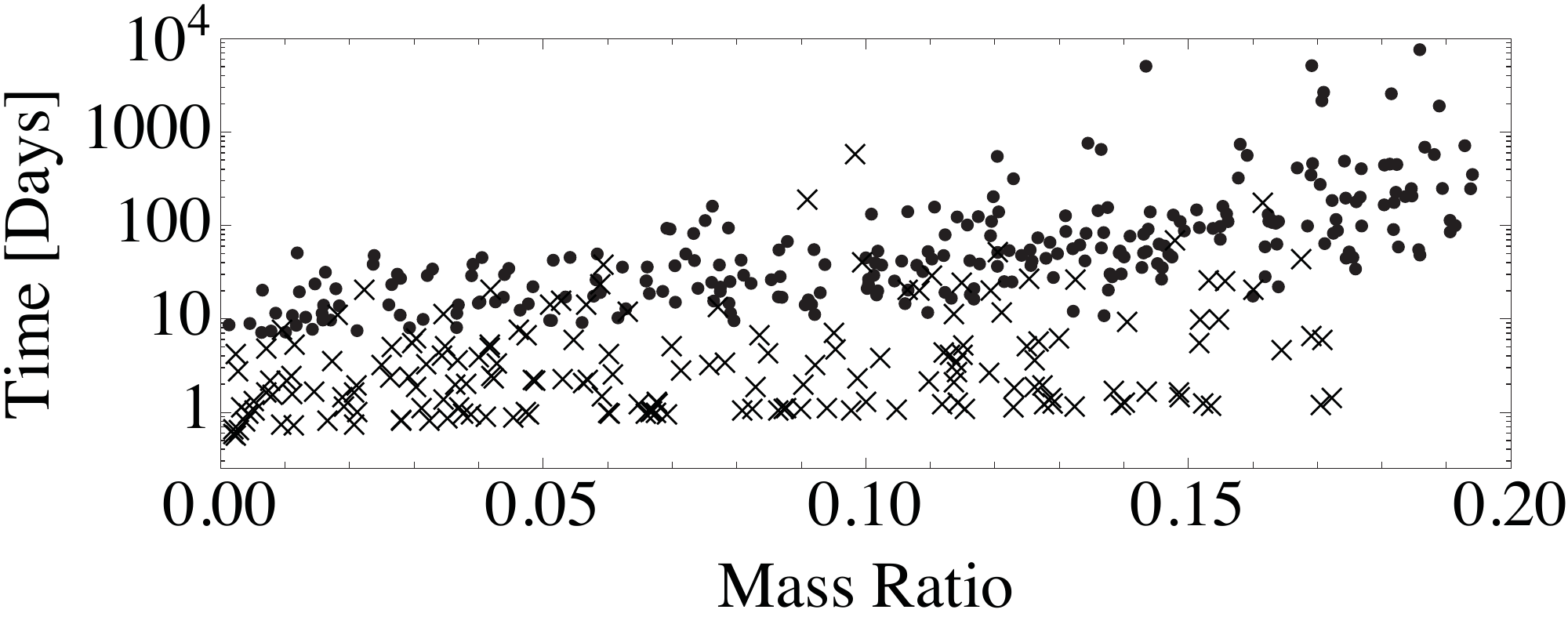}
\includegraphics[width=67mm]{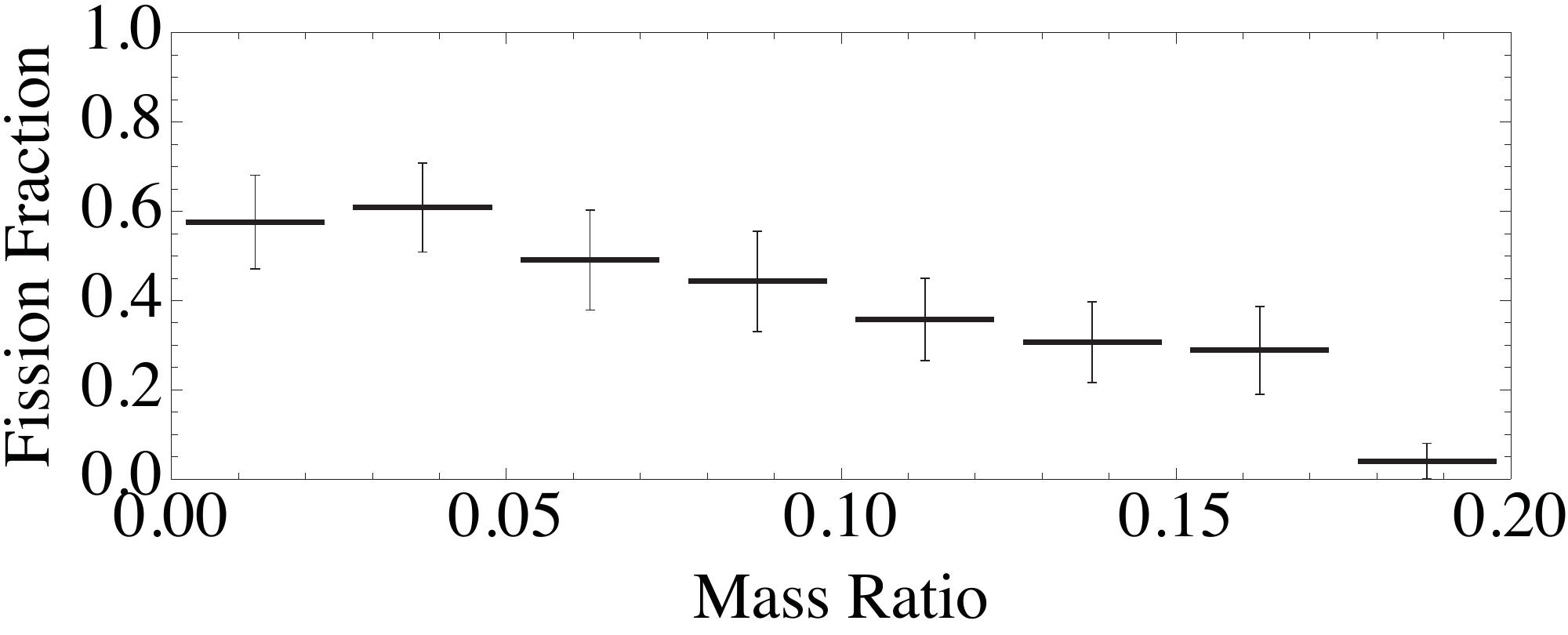}
\includegraphics[width=67mm]{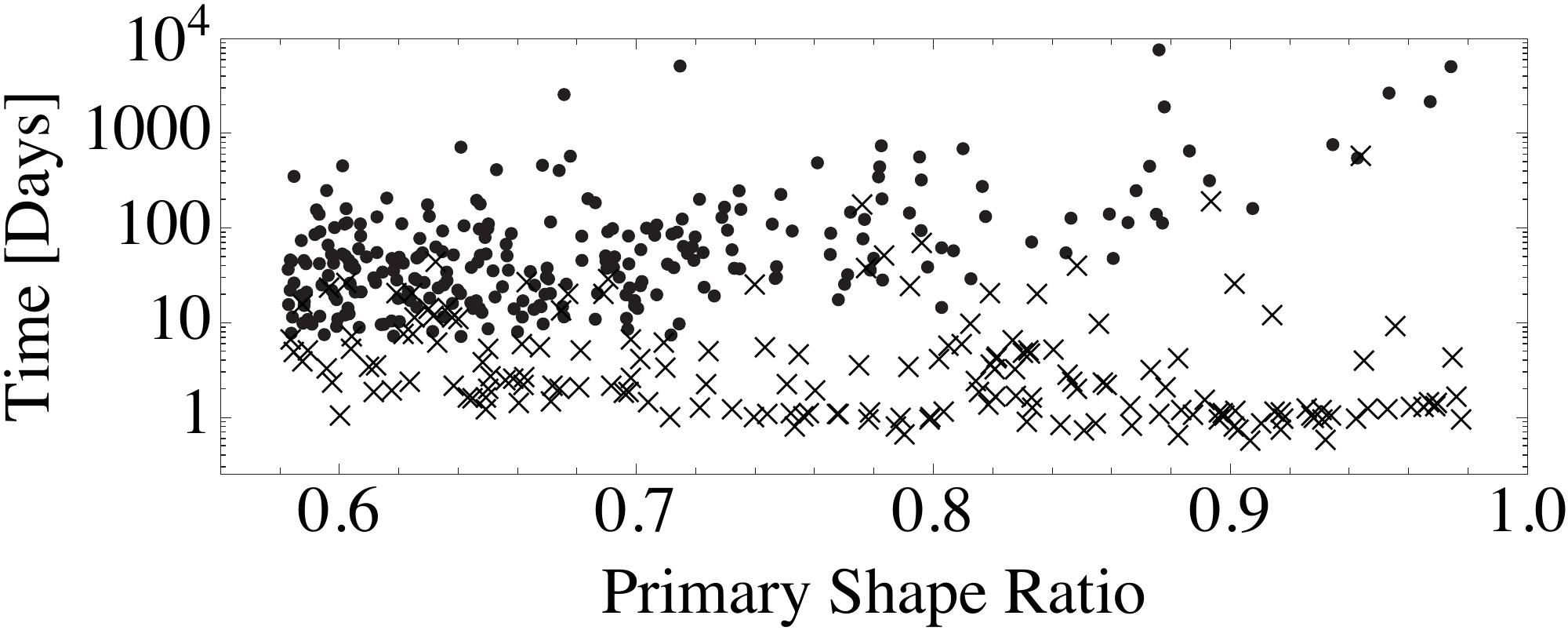}
\includegraphics[width=67mm]{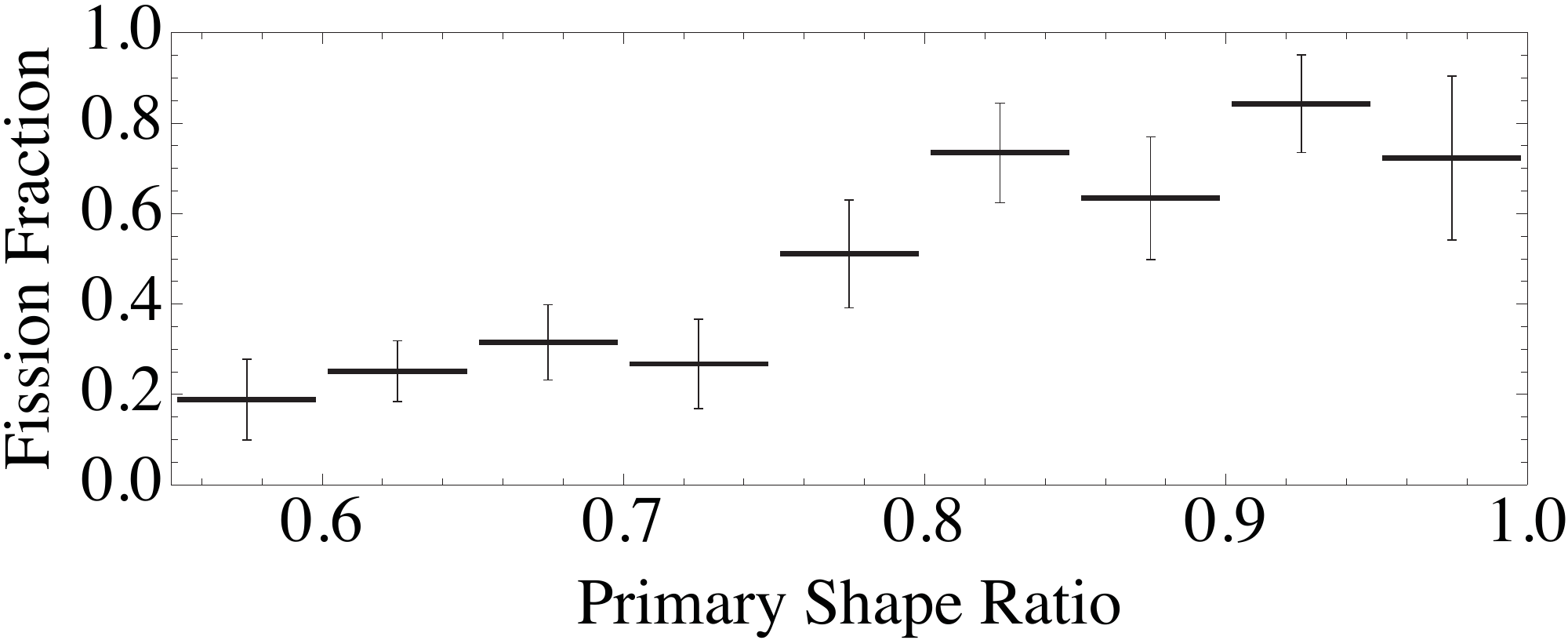}
\includegraphics[width=67mm]{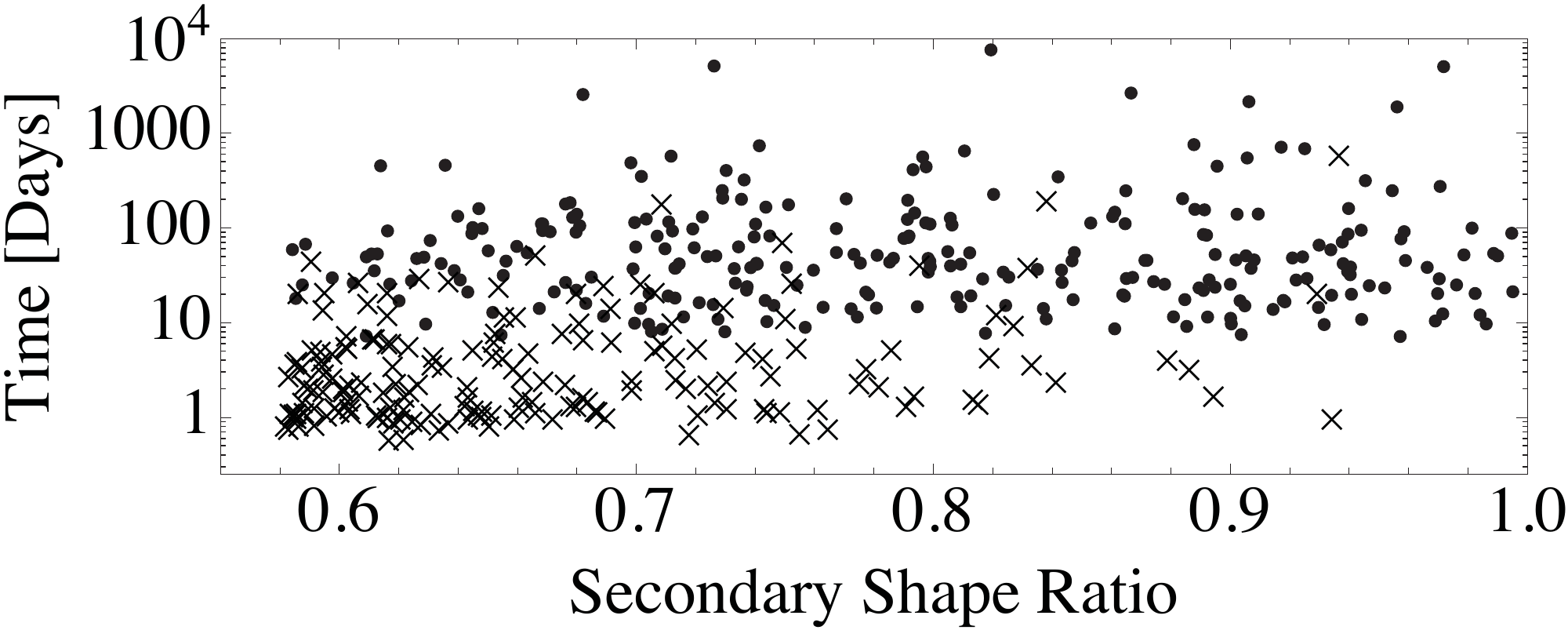}
\includegraphics[width=67mm]{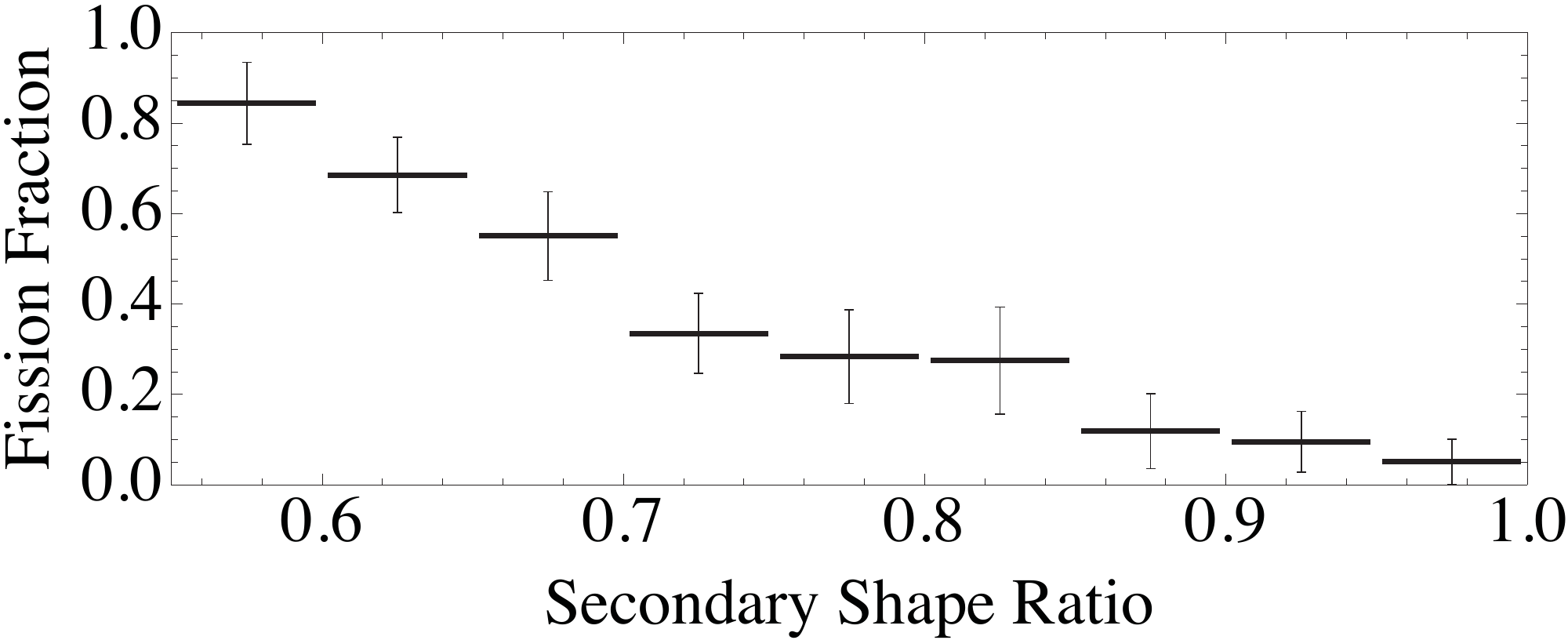}
\end{center}

\clearpage
Figure 9:
\begin{center}
\includegraphics[width= 90mm]{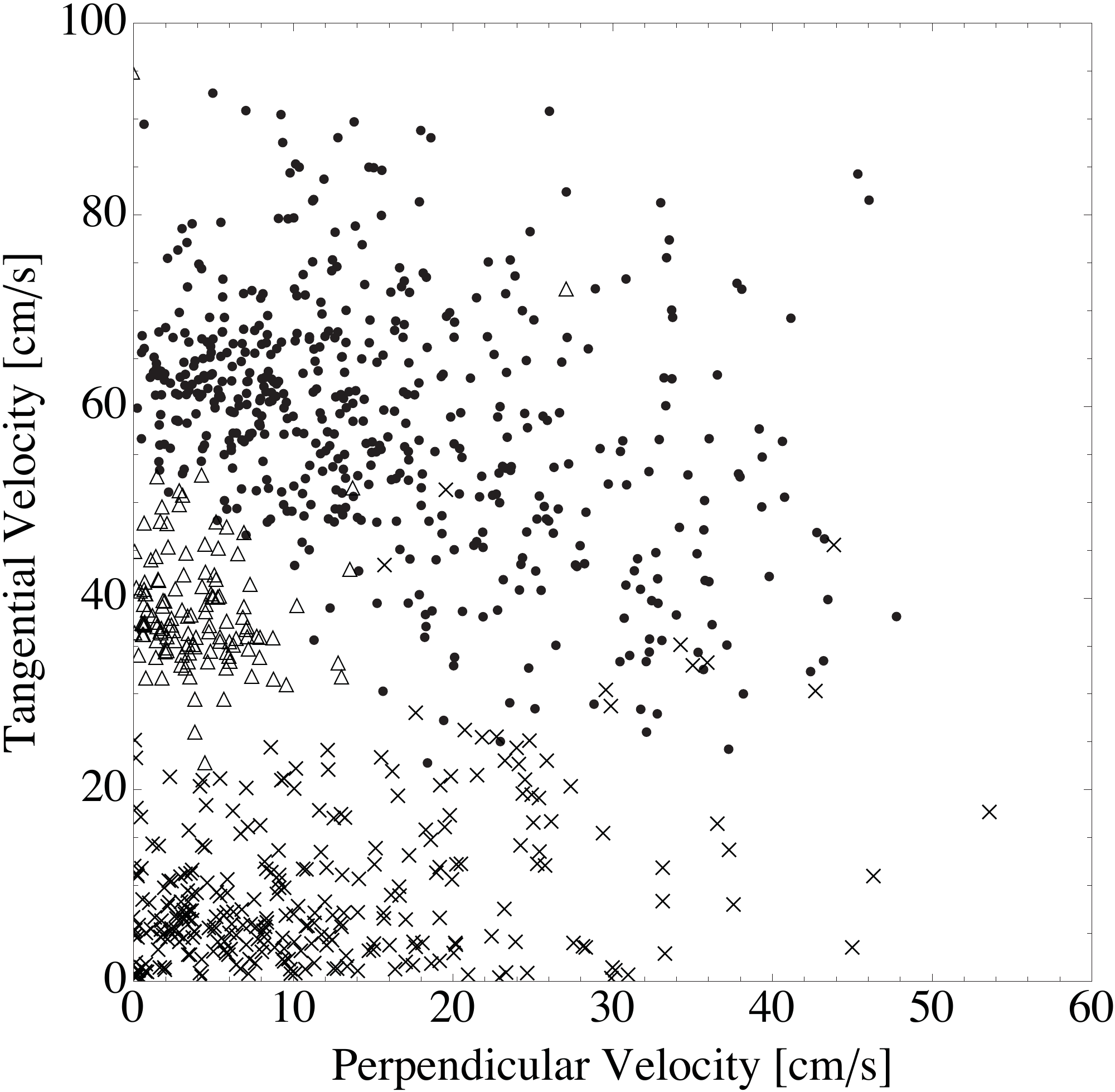} \\
\end{center}

\clearpage
Figure 10:
\begin{center}
\includegraphics[width= 90mm]{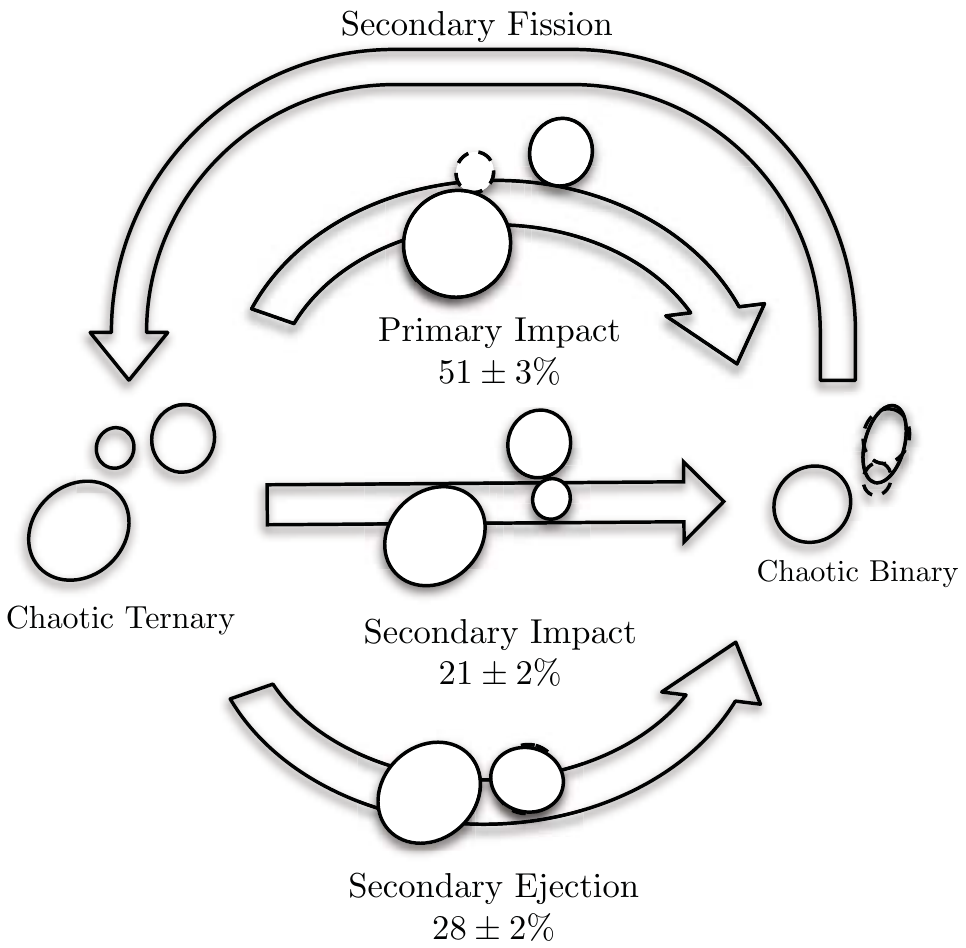}
\end{center}

\clearpage
Figure 11:
\begin{center}
\includegraphics[width= 90mm]{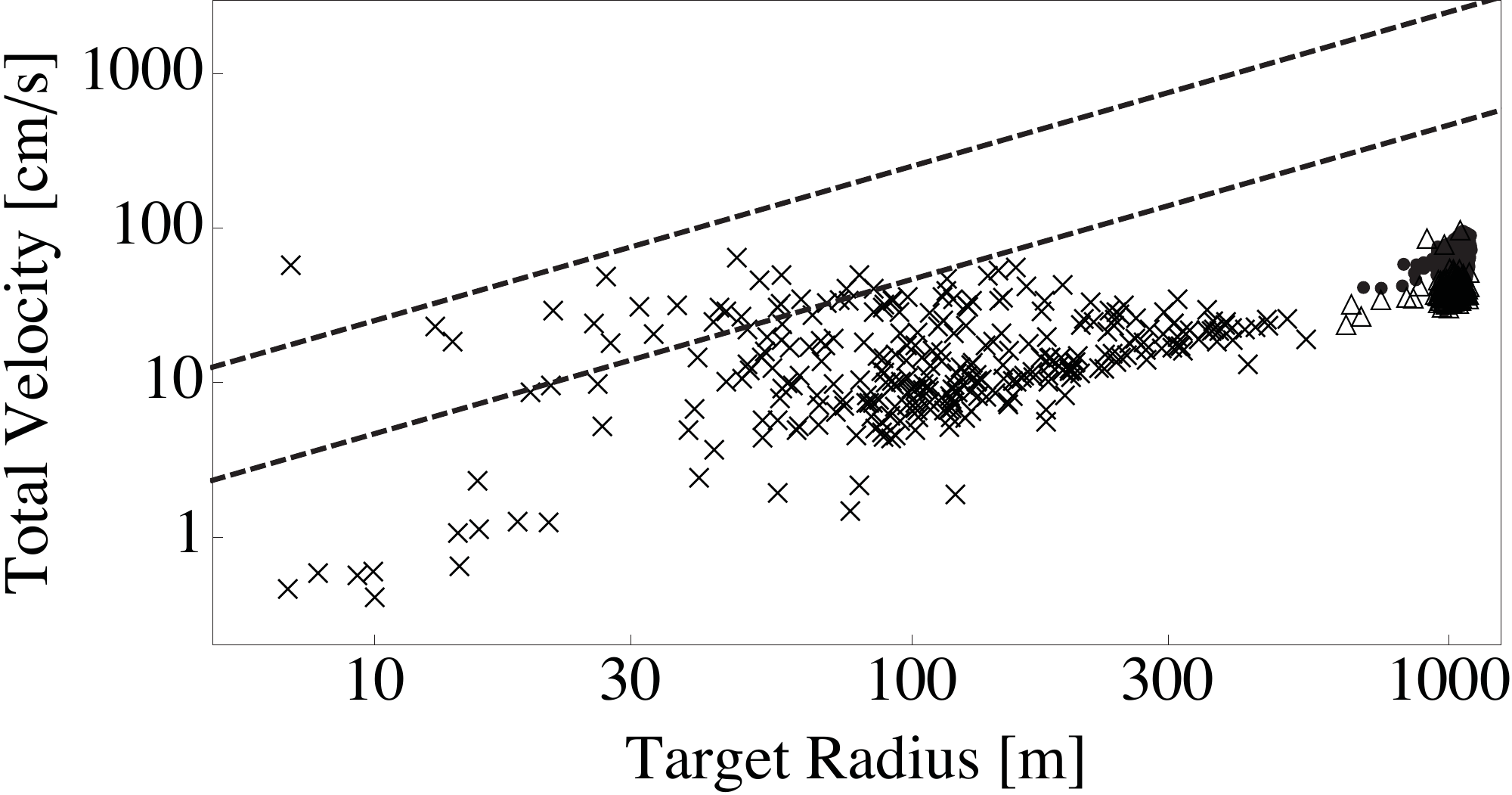} \\
\includegraphics[width= 90mm]{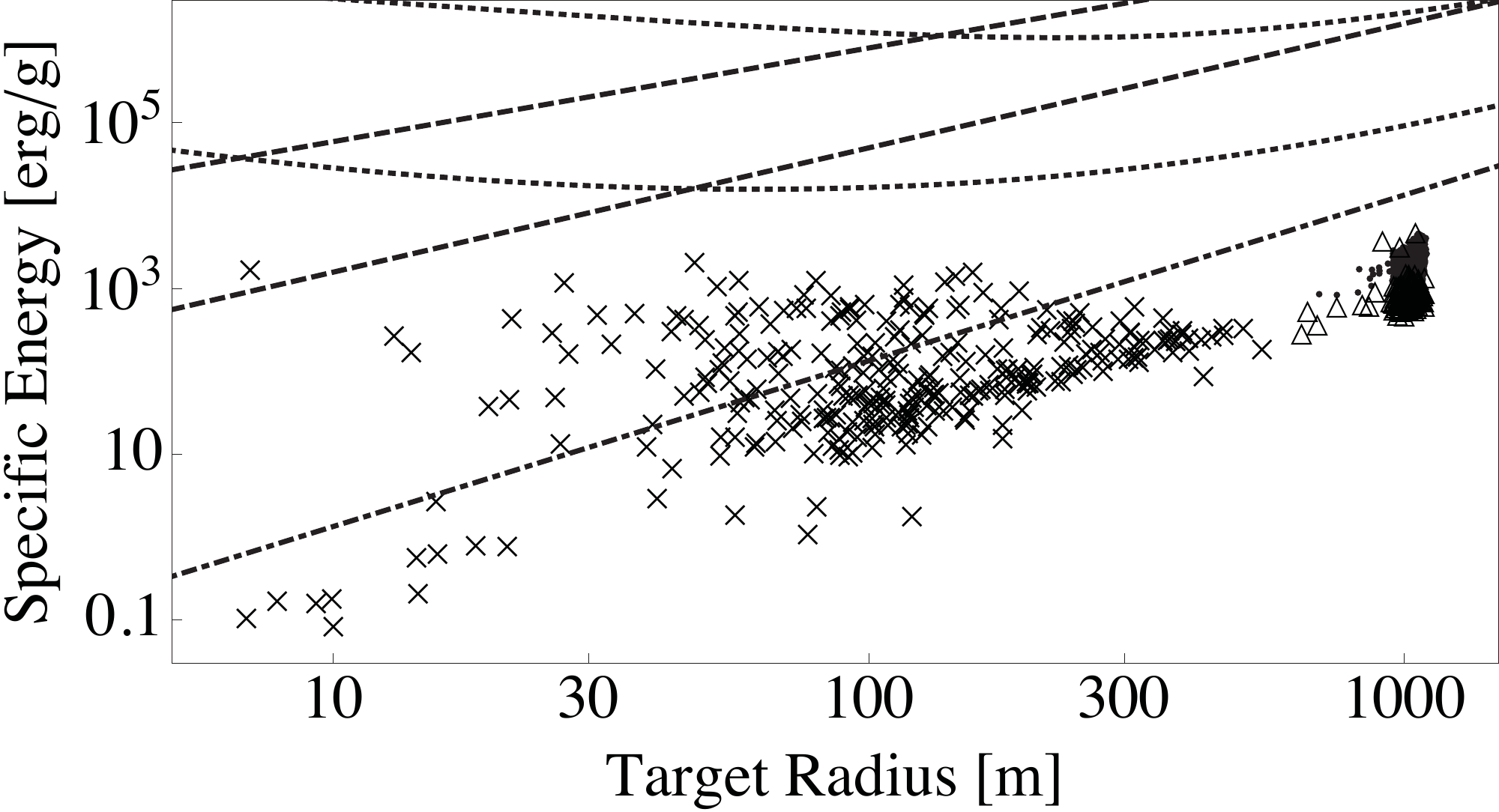}
\end{center}

\clearpage
Figure 12:
\begin{center}
\includegraphics[width=90mm]{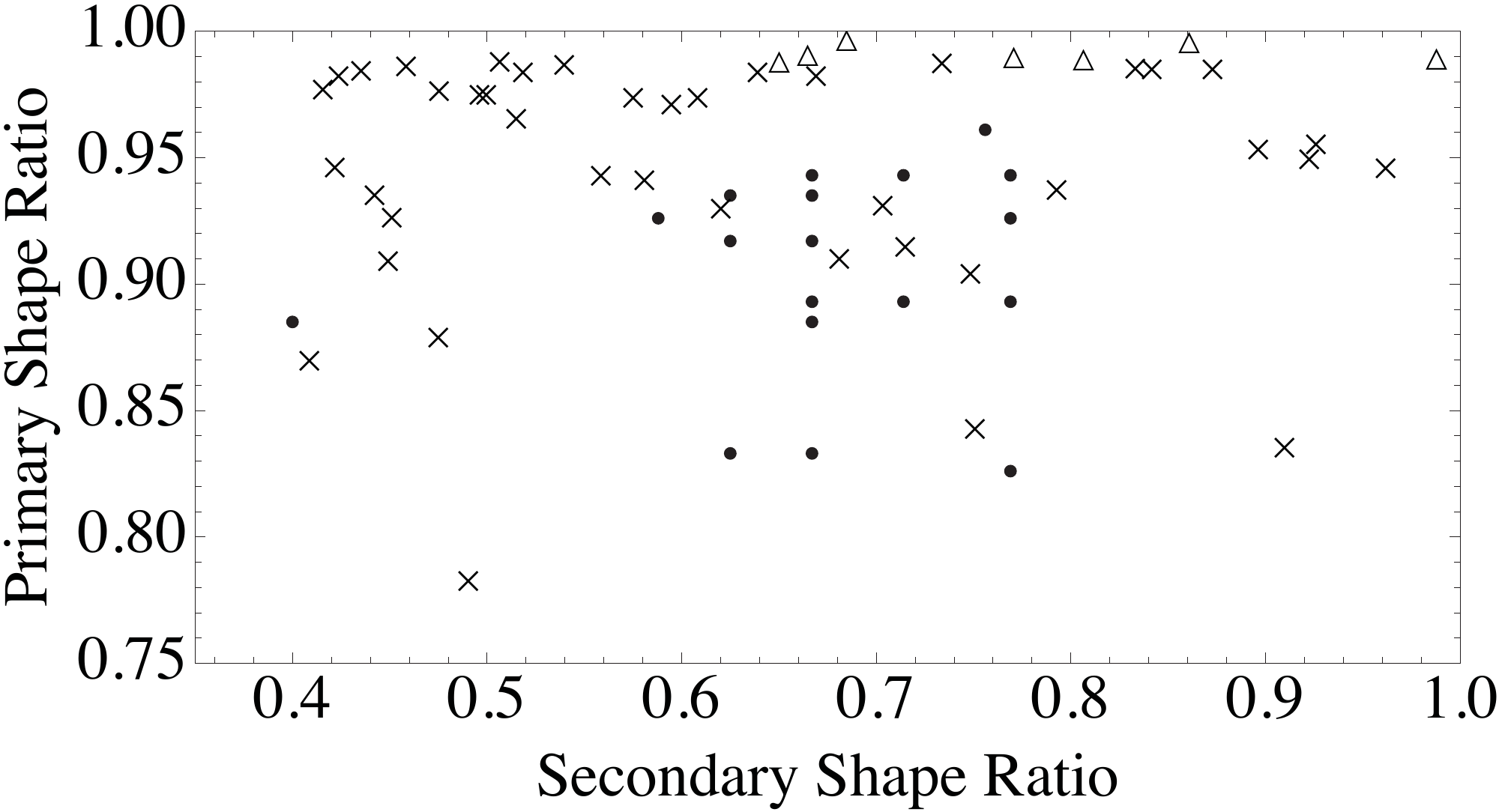}
\end{center}

\clearpage
Figure 13:
\begin{center}
\includegraphics[width=90mm]{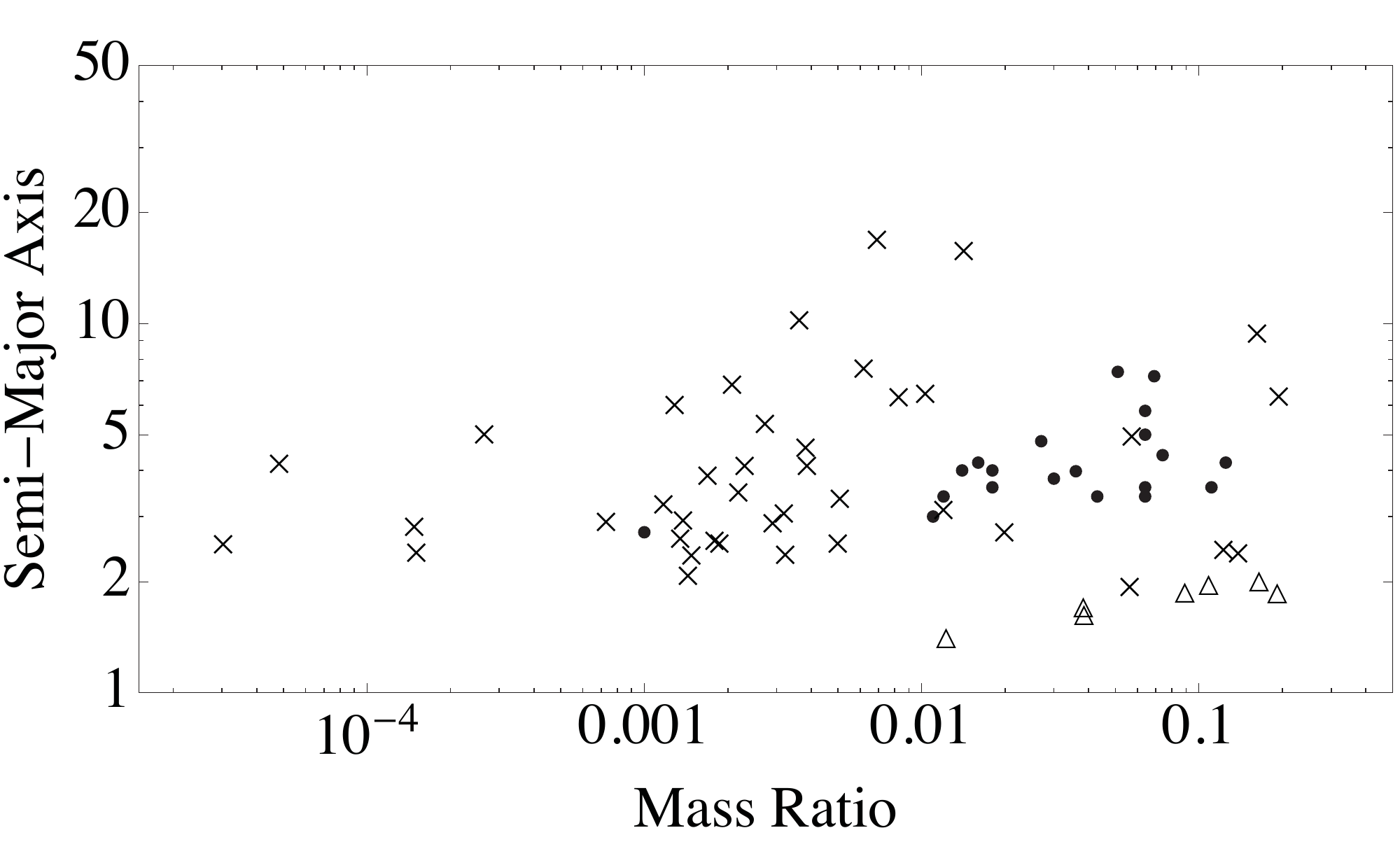} \\
\includegraphics[width=90mm]{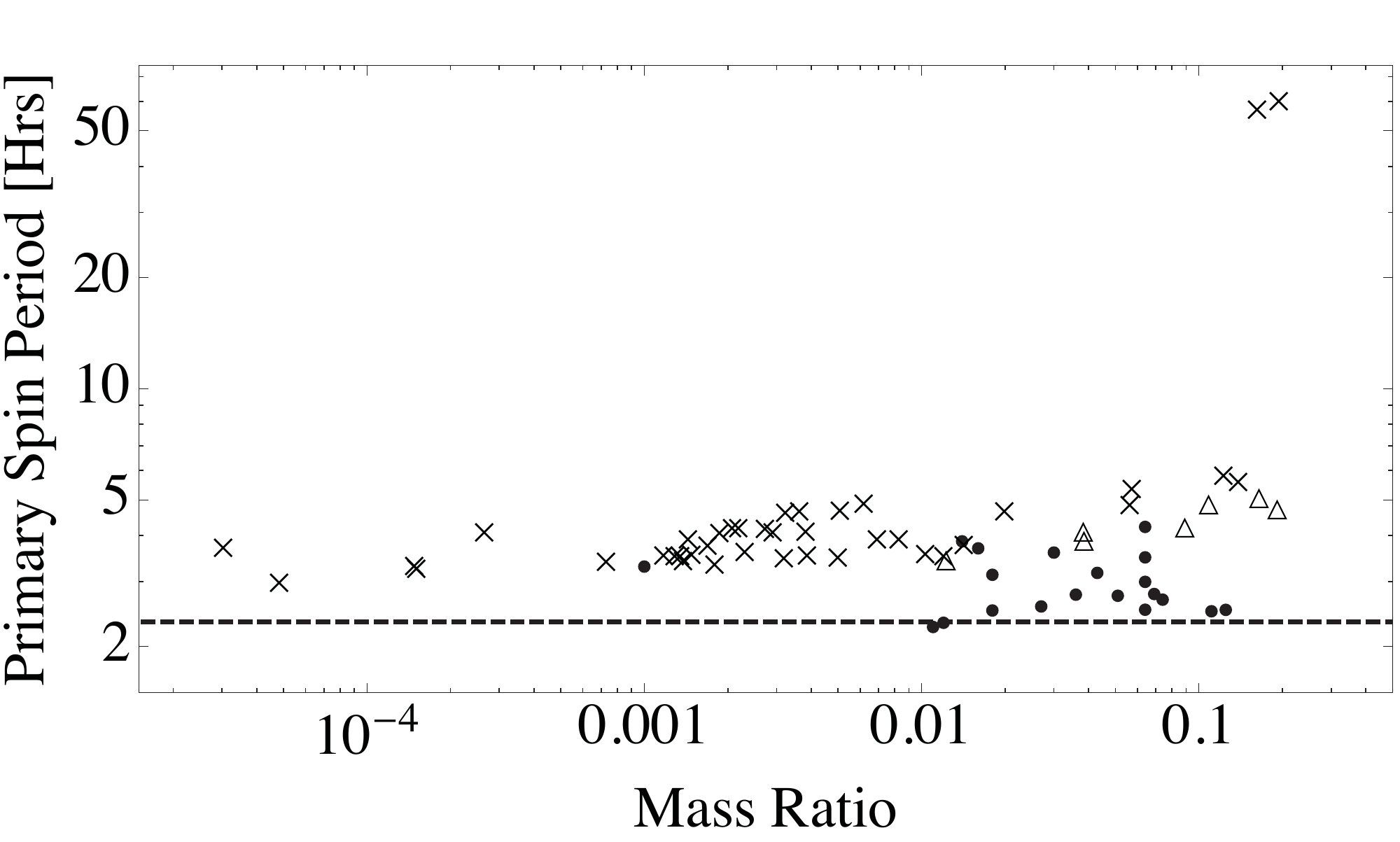} \\
\includegraphics[width=90mm]{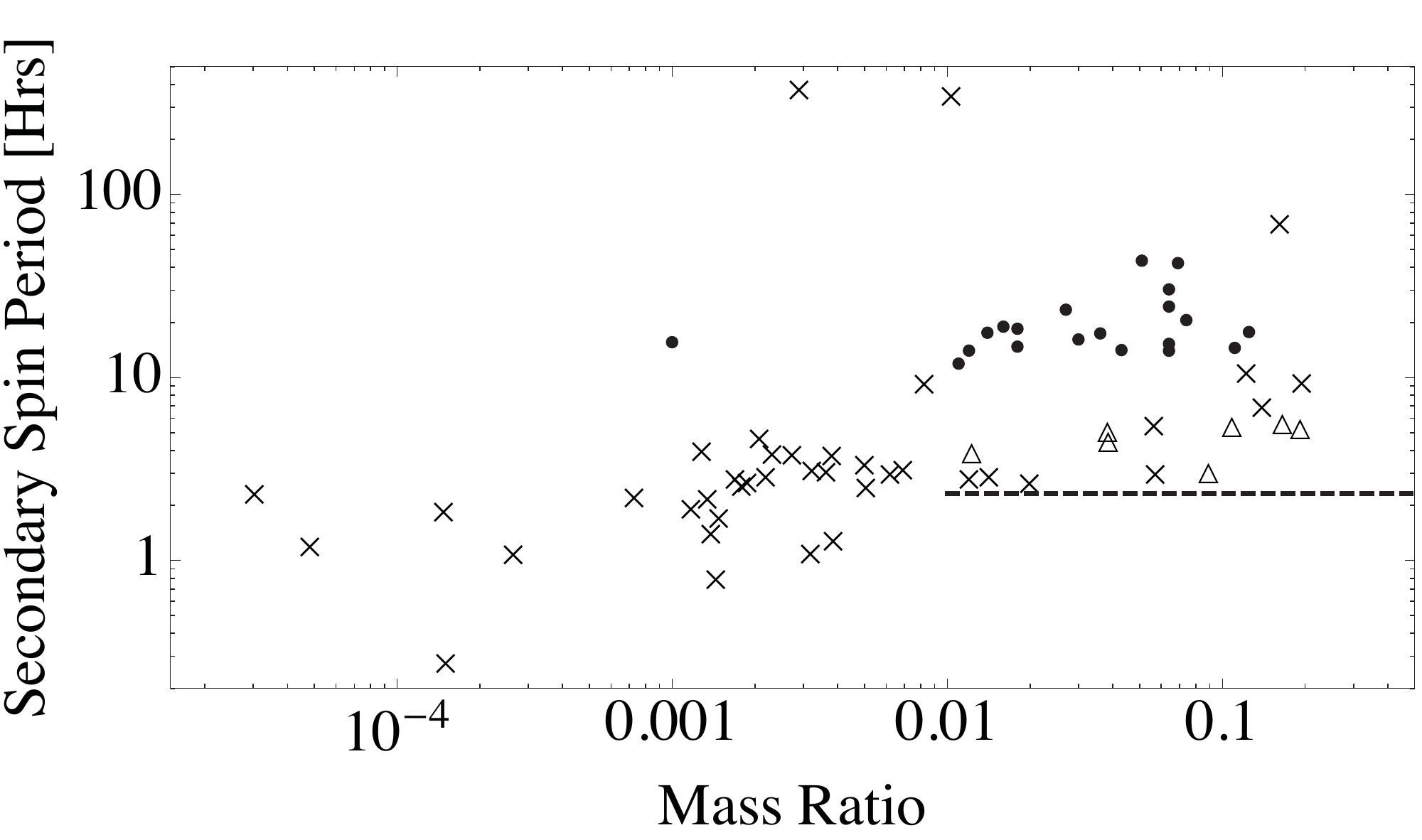}
\end{center}

\clearpage
Figure 14:
\begin{center}
\includegraphics[width= 90mm]{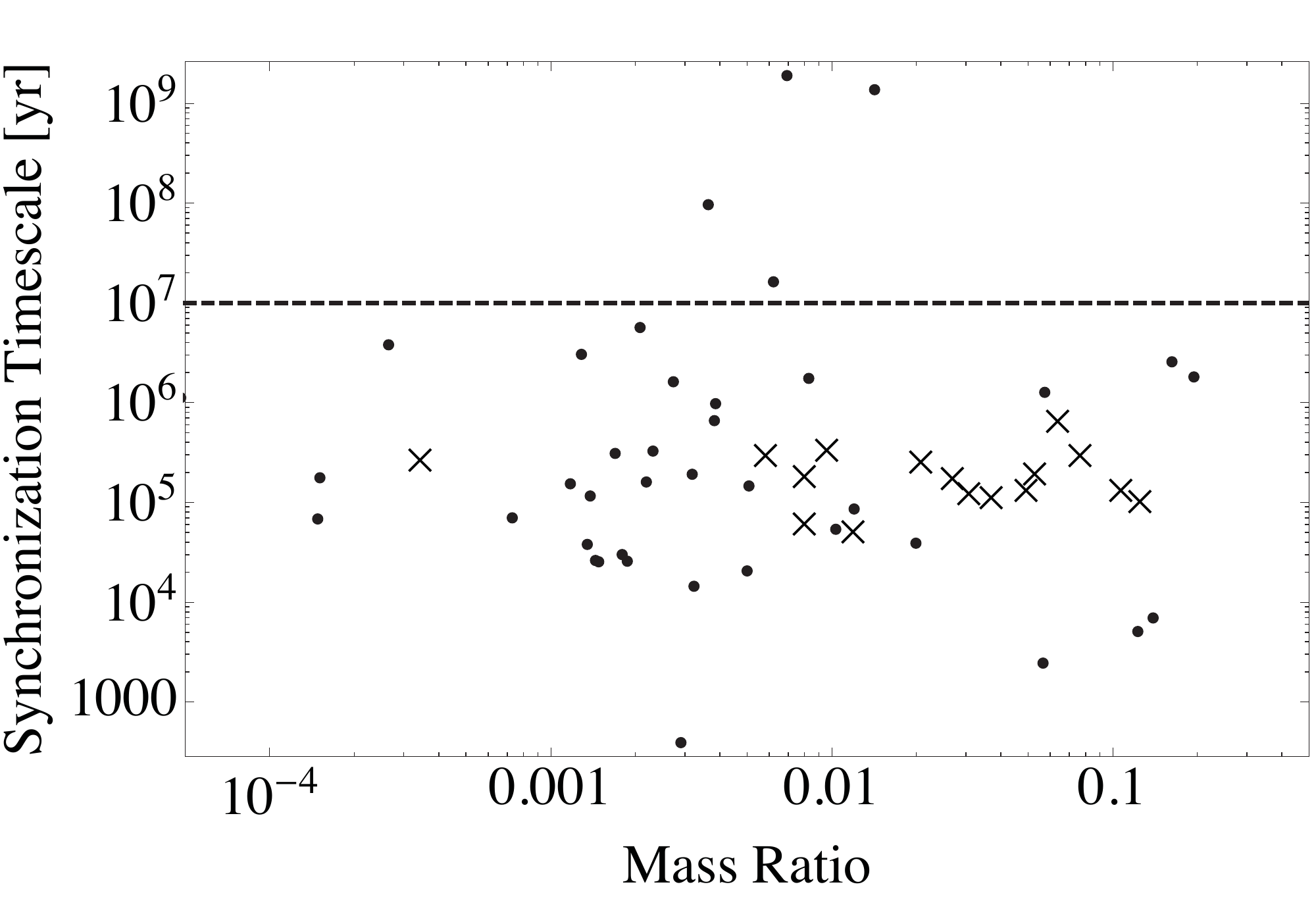}
\end{center}

\clearpage
Figure 15:
\begin{center}
\includegraphics[width= 90mm]{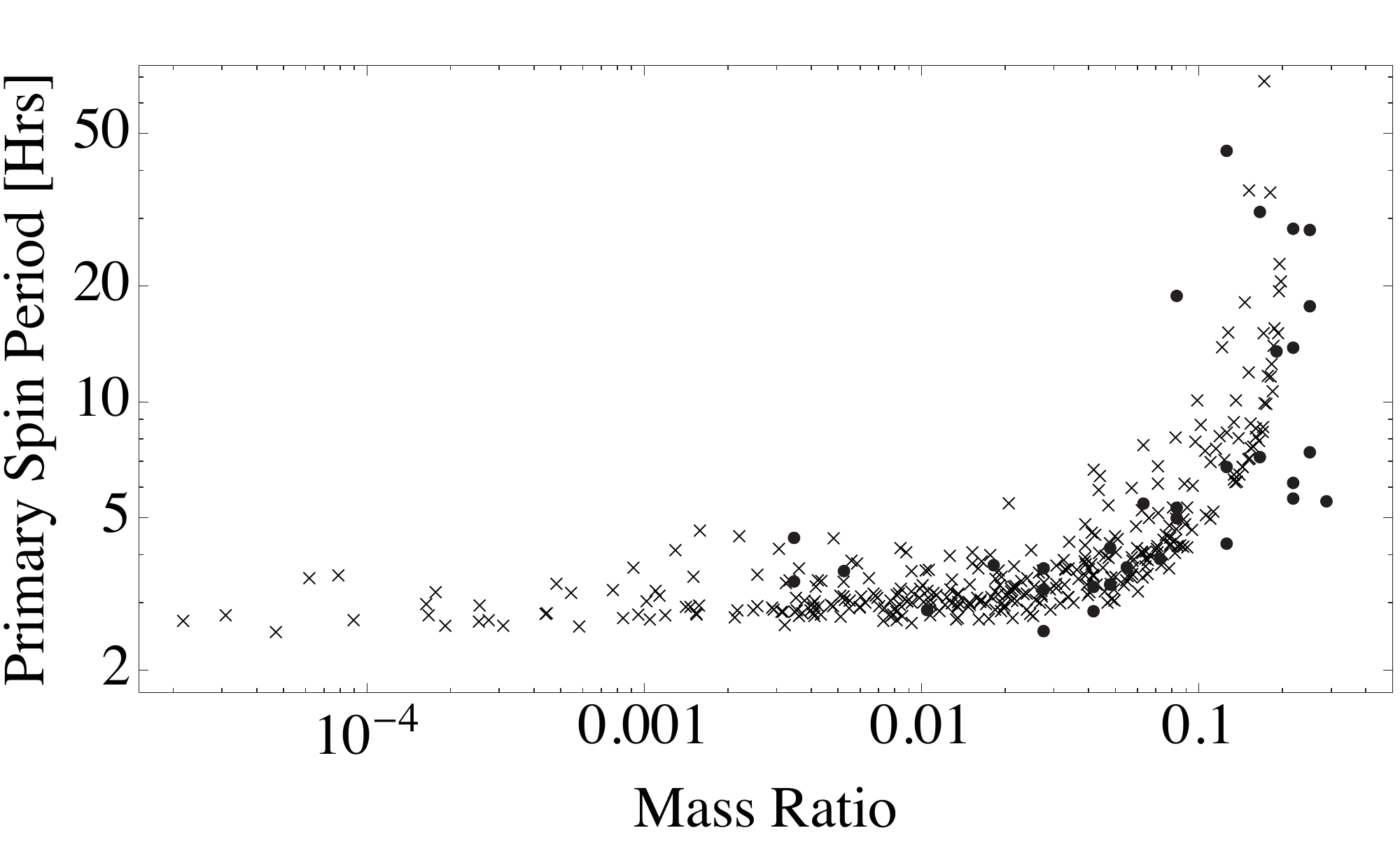}
\end{center}

\clearpage
Figure A.16:
\begin{center}
\includegraphics[width= 50mm]{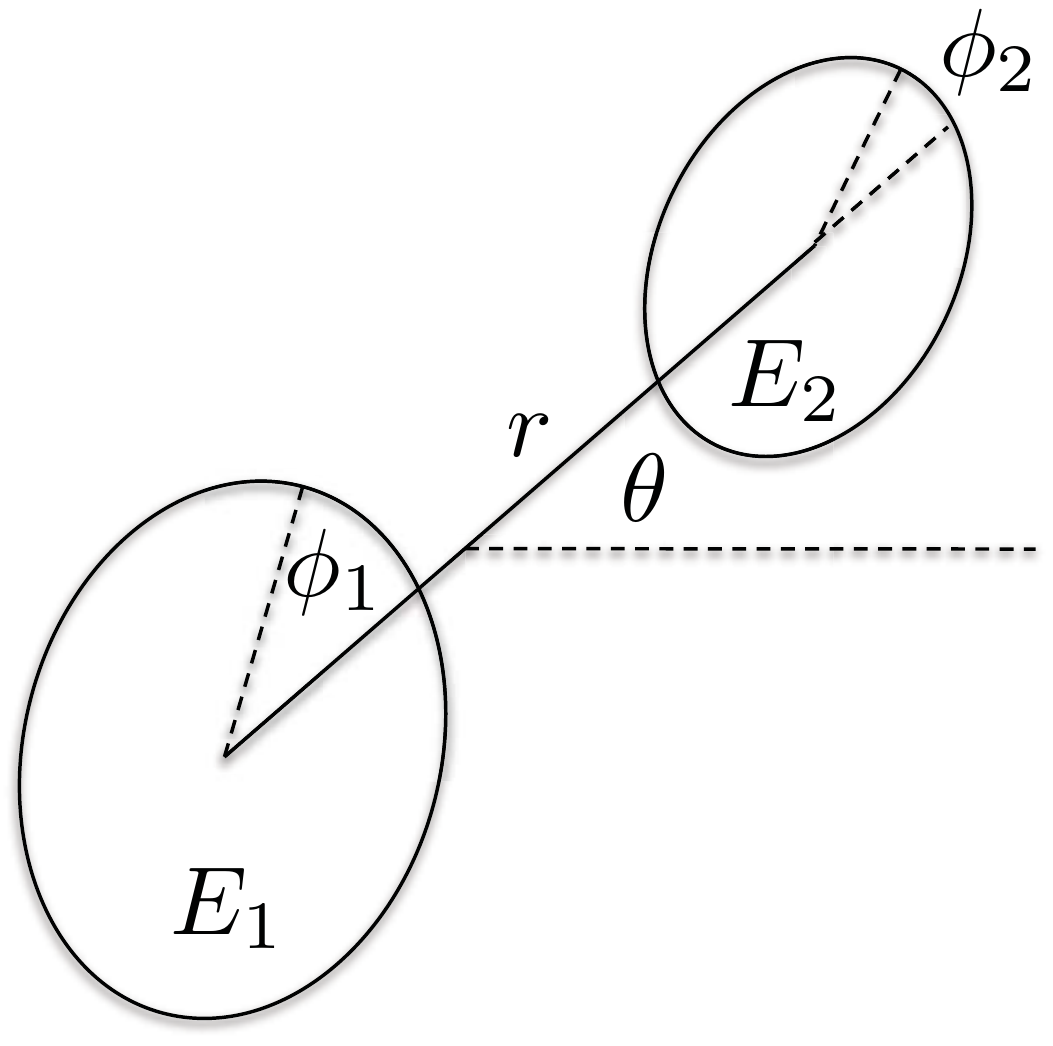}
\label{fig:TwoBodyCoordinates}
\end{center}

\clearpage
Figure B.17:
\begin{center}
\includegraphics[width= 50mm]{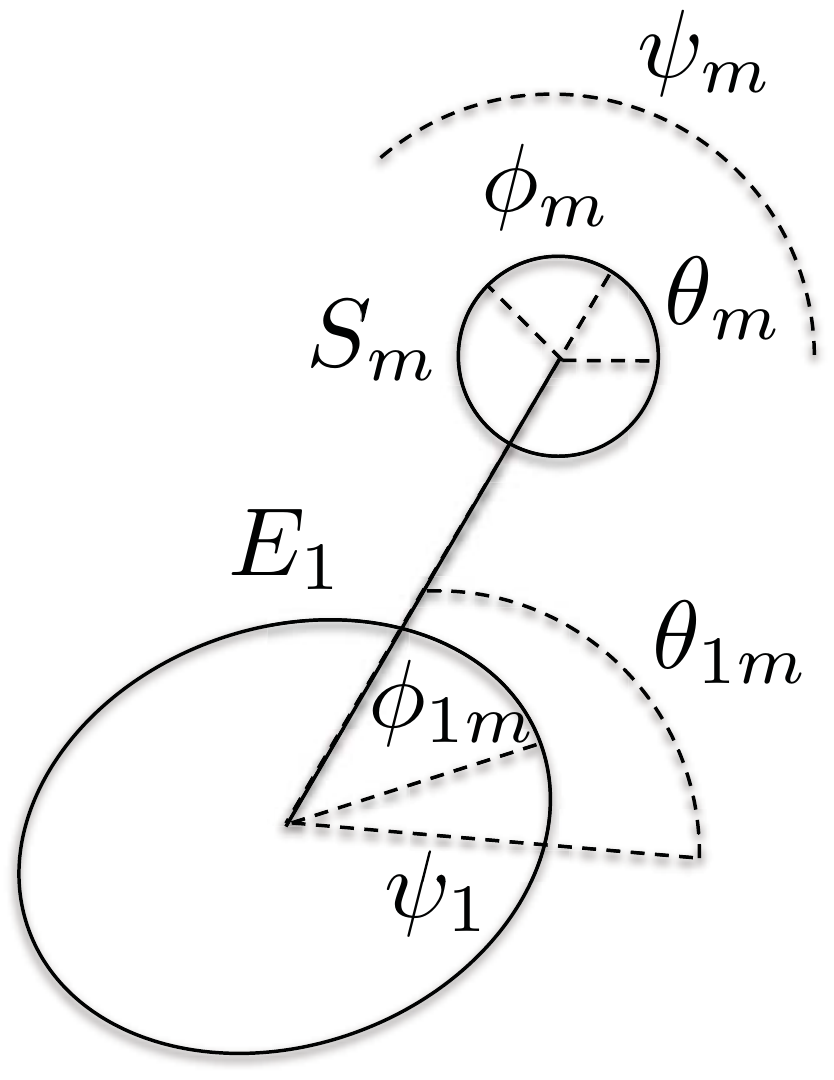}
\label{fig:ThreeBodyCoordinates}
\end{center}

\clearpage
Figure B.18:
\begin{center}
\includegraphics[width= 50mm]{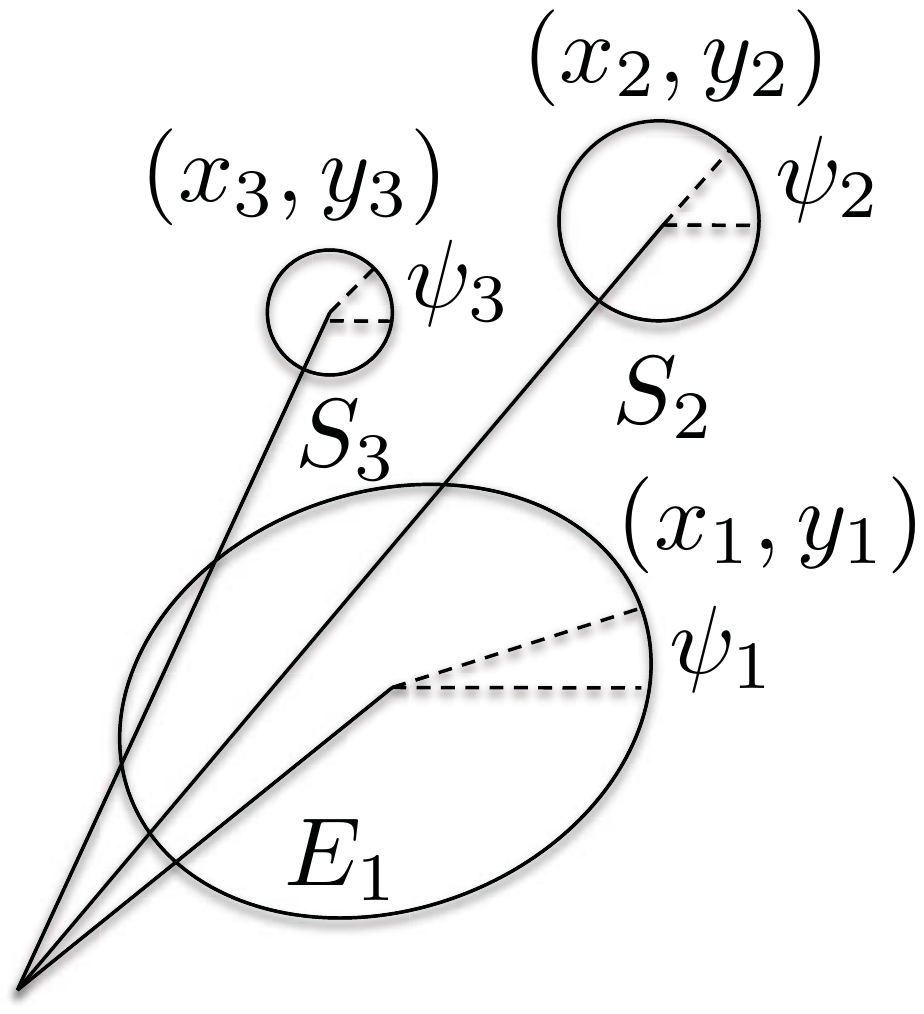}
\end{center}

\clearpage
Figure D.19:
\begin{center}
\includegraphics[width=90mm]{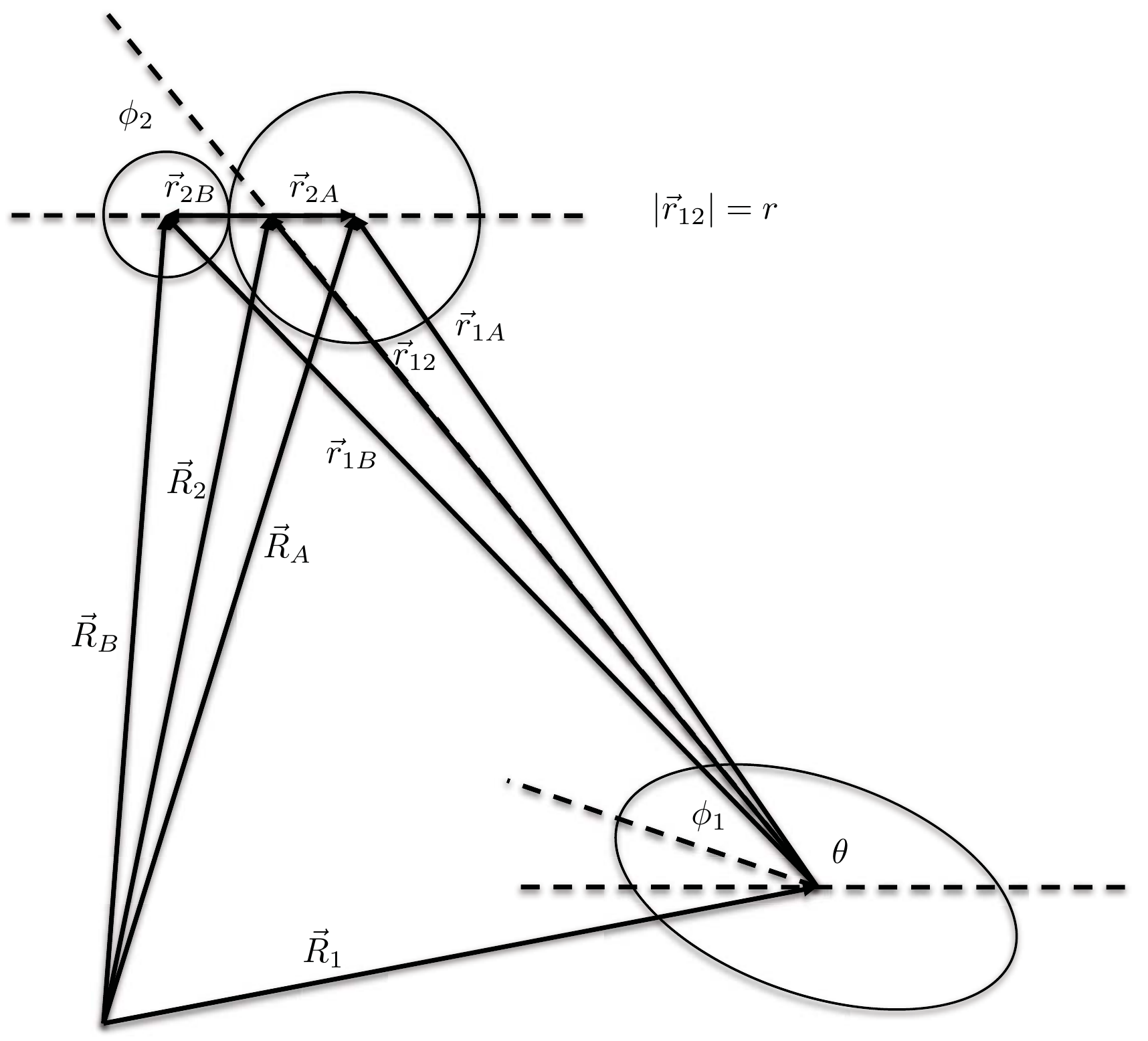}
\end{center}

\end{document}